\documentclass[structabstract]{aa} 
\usepackage{graphicx}
\usepackage{txfonts}
\usepackage{natbib}
\usepackage{color}
\begin{document}
   \title{Long Period Variables in the Large Magellanic Cloud from the EROS-2 survey \protect\footnotemark[1]}

   \author{M. Spano\inst{1}
          \and
          N. Mowlavi\inst{1,}\inst{2}
           \and
          L. Eyer\inst{1}
           \and
          G. Burki\inst{1}
             \and
          J.-B. Marquette\inst{3,}\inst{4}
             \and
          I. Lecoeur-Ta\"ibi\inst{2}
           \and
          P. Tisserand\inst{5,}\inst{6}
          }

   \institute{Observatoire de Gen\`eve, Universit\'e de Gen\`eve, 51 Chemin des Maillettes, 1290 Sauverny, Switzerland\\
                 \email{maxime.spano@unige.ch}
   	\and ISDC, Observatoire de Gen\`eve, Universit\'e de Gen\`eve, 1290 Versoix, Switzerland
   	\and UPMC Universit\'e Paris 06, UMR7095, Institut d'Astrophysique de Paris, F-75014, Paris, France
	\and CNRS, UMR7095, Institut d'Astrophysique de Paris, F-75014, Paris, France
         \and Research School of Astronomy \& Astrophysics, Mount Stromlo Observatory, Cotter Road, Weston ACT 2611, Australia
         \and CEA, DSM, DAPNIA, Centre dՃtudes de Saclay, 91191 Gif-sur-Yvette Cedex, France\\
          }

   \date{Received: May 20, 2011; accepted: September 19, 2011}

% \abstract{}{}{}{}{} 
% 5 {} token are mandatory
 
  \abstract
   {The EROS-2 survey has produced a database of millions of time series from stars monitored for more than six years, allowing us to classify some of their sources into different variable star types. Among the so-called, long period variables (LPVs), which are known to follow particular sequences in the period-luminosity diagram, we find long secondary period variables whose variability origin remains a matter of debate.}
   {We analyze data for the 856\,864 variable stars that have been discovered in the Large Magellanic Cloud (LMC) that are present in the EROS-2 database, to detect, classify, and characterize LPVs.}
   {Our method for identifying LPVs is based on the statistical Abbe test. It investigates the regularity of the light curve with respect to the survey duration in order to extract candidates with long-term variability. The period search is performed using Deeming, Lomb-Scargle, and generalized Lomb-Scargle methods, combined with a Fourier series fit.  Color-magnitude, period-magnitude, and period-amplitude diagrams are used to characterize our candidates.}
   {We present a catalog of 43\,551 LPV candidates for the Large Magellanic Cloud. For each of them, we provide up to five periods, a mean magnitude in EROS-2, 2MASS, and \textit{Spitzer} bands, $B_E-R_E$ color, $R_E$ amplitude, and spectral type. We use infrared data to distinguish between RGB, O-rich, C-rich, and extreme AGB stars. Properties of our LPV candidates are investigated by analyzing period-luminosity and period-amplitude diagrams.}
   {}

  \titlerunning{Long Period Variables in the Large Magellanic Cloud from EROS-2}
   \maketitle
%

%________________________________________________________________________________________
\section{Introduction}
\footnotetext[1]{Full Tables 3 is only available in electronic form at the CDS via anonymous ftp to cdsarc.u-strasbg.fr
(130.79.128.5) or via http://cdsweb.u-strasbg.fr/cgi-bin/qcat?J/A+A/}
Long period variables (LPVs) are red giant stars whose brightness varies on timescales from weeks to years.
They are either mono- or multi-periodic, and include variable stars from the variability types formerly classified as Mira, semi-regular, and long secondary period (LSP) variables.
Among the distinctive characteristics of LPVs are the period-luminosity (PL) relations that they follow.
A relation between the period and the absolute magnitude for Mira variables was first proposed by \citet{Gerasimovic28}. Thanks to $J$, $H$, and $K$ band near-infrared data, \citet{GlassLloyd81} were able to refine this relation that until then had displayed large scatter, as observations had been made using visual magnitude. \citet{WoodSebo96} found a sequence in the ($K$, log $P$) plane that is parallel to the Mira PL relation at shorter periods and compatible with a first overtone pulsator population.

The advent of large-scale surveys such as MACHO and OGLE has opened a new dimension in the study in general of variable stars, and in particular of LPVs.
Initially devoted to the search of microlensing events, these surveys provided large databases that made the statistical study of LPVs highly significant.
\citet{Woodetal99} identified five sequences in the PL diagram, using MACHO data \citep{Alcocketal92} of the Large Magellanic Cloud (LMC). 
They were labelled A, B, C, E, and D with increasing periods. The previously known PL relation of Mira stars falls on sequence C and corresponds to the radial fundamental mode of pulsation.
Sequences A and B were identified to be third, second, or first overtone pulsators and comprise semi-regular variables.
\citet{Itaetal04}, using OGLE-II \citep{Udalski97} and SIRIUS \citep{Nagashima99} surveys, showed that sequence B could be separated into two sequences, called B and C'. Using pulsation models from \citet{WoodSebo96} they concluded that sequence C' is composed of "Mira-like" variables pulsating in their first overtone, complementing thereby the Mira fundamental mode pulsators of sequence C. However, the theoretical periods for the second and third overtones do not fit their observed sequences B and A well. Therefore, models for these two sequences are still to be found.

 Theory predicts that fundamental radial mode pulsators have the longest periods, though sequences E and D appear to be at longer periods than sequence C.
Since the study of \citet{Woodetal99}, sequence D is known to contain LSP variables. These stars were characterized as having two periods, with a period ratio of the longer to the shorter period of 9$\pm$4, the longer period belonging to sequence D and the shorter, usually, to sequence B. Using OGLE-II and OGLE-III \citep{Udalskietal08} photometry, \citet{Soszynskietal04} suggested that sequence E may be the prolongation of sequence D at lower magnitude. As sequence E was found to be populated by ellipsoidal binaries, the shift of sequence E to periods shorter than those of sequence D would be due to the binary nature of the objects. The orbital periods of those stars should actually be twice as long as found, placing them on line with sequence D, which, as a consequence, would also contain binaries. However, \citet{Nichollsetal10} excluded the hypothesis of binary systems for LSPs. Hence, the origin of variability for LSPs is still debated \citep{Nichollsetal09, Woodetal09, Nieetal10}.

The analysis of the data from other large-scale surveys provided further information about LPVs.
While sequences A to D published by \citet{Woodetal99} was assumed to consist of only asymptotic giant branch (AGB) stars, \citet{KissBedding03} demonstrated that the sequences extended below the tip of the red giant branch using OGLE-II and 2MASS \citep{Skrutskieetal97} data, establishing the presence of red giant branch (RGB) pulsators as well.

\begin{table*}
\caption{Observational properties of the EROS-2 LMC survey.}             % title of Table
\label{table1}      % is used to refer this table in the text
\centering                          % used for centering table
\begin{tabular}{c r r r c c c c}        % centered columns (4 columns)
\hline
Target & Number    & Total number   & Number of      & \multicolumn{2}{c}{Mean number obs. per star} & \multicolumn{2}{c}{Mean duration (days)}\\    
  ~    & of fields & of stars & variable stars & $R_E$ & $B_E$ & $R_E$ & $B_E$ \\
\hline                        % inserts single horizontal line
   LMC                  & 88 & 28\,800\,222 & 856\,864 & 389 & 477 & 2\,002 & 2\,324\\
\hline                                   %inserts single line
\end{tabular}
\end{table*}

In this paper, we search for LPVs in the LMC and complie a catalog, from the database of variable stars provided by the EROS-2 survey.
The EROS-2 survey was initiated in 1996, and during almost seven years monitored the Magellanic Clouds, the Galactic bulge, and four fields in the Galactic disk.
Despite the rich database gathered during its seven years of operation called the EROS-2 catalog of variable stars that records, for example, 856\,864 variables in the LMC fields (Marquette et al., 2011  in prep.), the survey remains to date under-exploited in the field of red-giant variable-star studies.

The catalog published in this paper comprises 43\,551 variable stars that we identify as LPV candidates in the EROS-2 survey of the LMC.
This survey is particularly well-suited as it covers a sky area of 88 deg$^2$ on the LMC, compared to 13.5 deg$^2$ for MACHO and 40 deg$^2$ for OGLE-III.

We start Sect.~\ref{Sect:theSurvey} by recalling the main characteristics of the EROS-2 survey of the LMC and of its variable star content.
The identification of LPV candidates of the LMC is then described in Sect.~\ref{Sect:Identification}. The computation of pulsation periods  and some characteristics of the selected LPV candidates are shown in Sect.~\ref{Sect:Periodsearch}.
 In Sect.~\ref{Sect:Properties}, we discuss their infrared, period-luminosity, and period-amplitude properties. Conclusions are finally presented in Sect.~\ref{Sect:Conclusions}.
%________________________________________________________________________________________
\section{The EROS-2 survey}
\label{Sect:theSurvey}
%----------------------------------------------------------------------------------------
\subsection{Presentation of the EROS-2 survey}

The EROS-2 survey operated from July 1996 to March 2003 with seven targets, namely the LMC, the Small Magellanic Cloud (SMC), the Galactic center, and fields around $\theta$ Mus, $\gamma$ Nor, $\gamma$ Sct, and $\beta$ Sct in the disk of the Galaxy.
Observations were made with the MARLY, a 1 m Ritchey-Chr\'etien telescope operating at La Silla Observatory in Chile.
Two cameras, with a pixel size of 0.6" and a field of view of 0.7x1.4 deg$^{2}$ on the sky, taken data simultaneously in the non-standard blue $B_E$ (420 -- 720 nm) and red $R_E$ (620 -- 920 nm) bands.

 The observation properties of the EROS-2 LMC survey are summarized in Table~\ref{table1}.
 Owing to some problems with the 'red' camera towards the end of the survey program, the time series in the $R$ band are shorter than in the $B$ band by about 300 days.
 Nevertheless, we use R band data for our analysis, since LPVs are red and hence benefit from a higher photometric precision in the $R$ band than in the $B$ band.
 Data pre-reduction, dealing with corrections for a flat-field, bias, and dark current, was directly made at the telescope.
 Data reduction and analysis were then made at the IN2P3 computing center in Lyon, France, with the dedicated programs package PEIDA \citep[see][]{Ansari96}.
 Our work is based on the resulting light curves provided by the EROS team.

A total of 87 million sources have been observed by EROS-2 over its whole mission, 4.6~\% of them being classified as variable (Marquette et al., 2011 in prep.), which correspond to a database of time series for more than 4 million sources.
One of the most populated target is the LMC, with a total of 28.8 million observed stars of which about 3~\% were detected as variable (cf. Table~\ref{table1}).

%----------------------------------------------------------------------------------------
\subsection{LMC variable stars in the EROS-2 database}
\begin{figure}
	\includegraphics[width=\columnwidth]{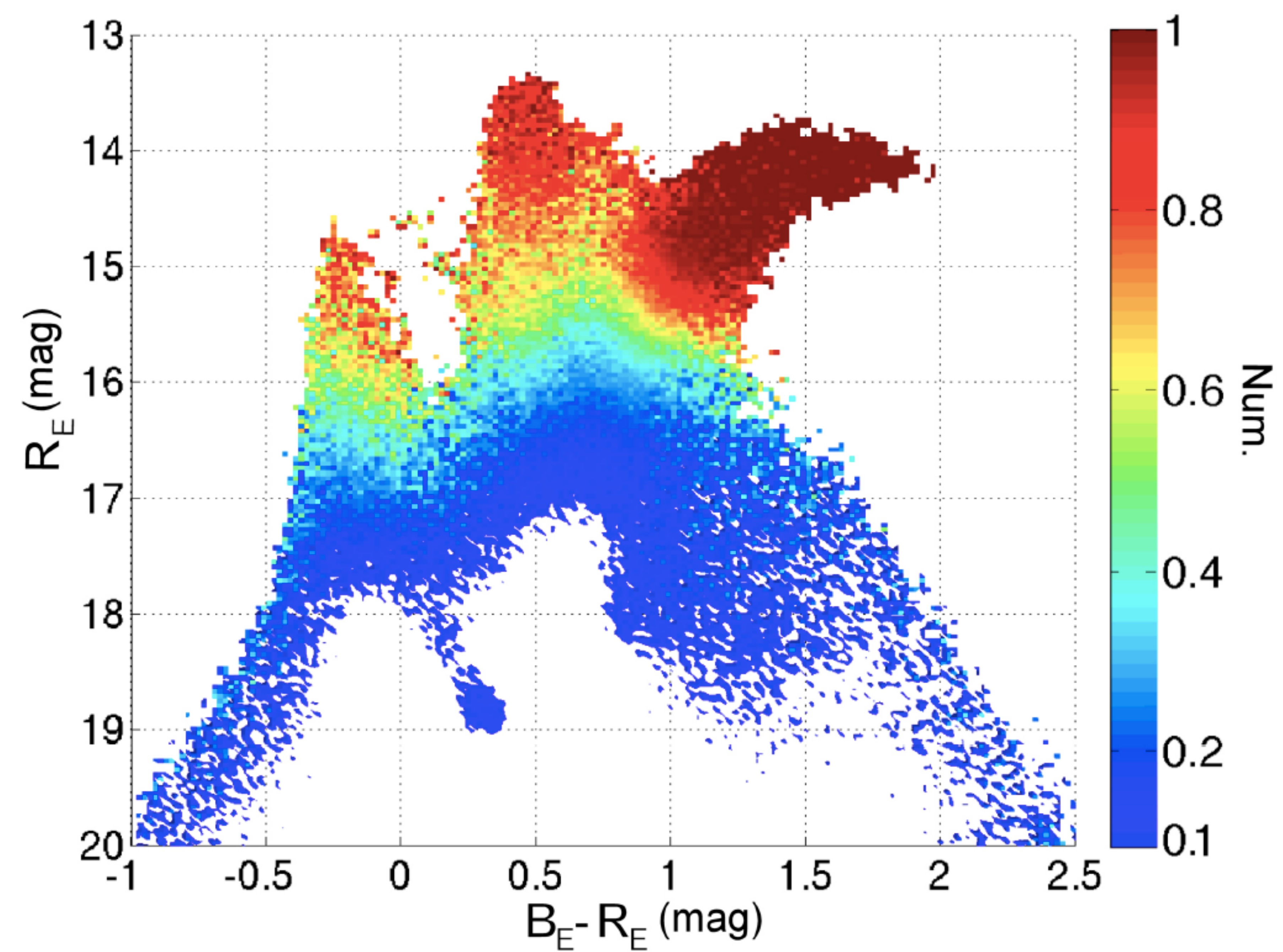}
   	\caption{EROS-2 ($B_E-R_E$, $R_E$) color-magnitude diagram of the ratio of the number of variable stars to the total number of stars.
	The color-magnitude diagram is divided into a matrix of 250x250 bins, each bin being 0.014x0.028 mag wide.
	The ratio is plotted according to a color-scale shown on the right of the figure.
	Only bins with at least ten stars and a ratio above 0.1 are plotted.  Color version of the figures in this paper are available in the online version.}
   	\label{CMDRatio}
\end{figure}

The ratio of the number of variable stars to the total number of stars across the color-magnitude diagram (CMD) is shown in Fig.~\ref{CMDRatio}. It displays the fraction of LMC stars if at least 10\% of them are variable in each bin. 
Four structures are clearly visible in this figure. 
The one around $B_E-R_E=-0.25$ mag and spreading from magnitudes $R_E=18$ mag down to $R_E=14.7$ mag, represents the main sequence and would contain, among others, eclipsing binaries and variables of the Be type.
The second one at $B_E-R_E=0.33$ mag and $R_E=18.8$ mag is the RR Lyrae clump. Despite their faint luminosities, they are identifiable thanks to their amplitude of variability of about magnitude.
Another vertical structure is visible at $B_E-R_E=0.4$ mag and from $R_E=17$ mag down to $R_E=13.5$ mag.
This structure contains Cepheids, but might also be populated by some foreground stars from our Galaxy.
The last structure is the AGB extending to the redder and brighter part relative to the tip of the RGB visible at $B_E-R_E=1.1$ mag and $R_E=15$ mag. Almost all AGB stars are variable, and thus potential LPVs.
 The lack of points at the faintest magnitudes does not necessarily reflect an absence of variable stars, but rather an observational bias because the variability of fainter stars is more difficult to detect at these magnitudes owing to the larger photometric errors.
%----------------------------------------------------------------------------------------
\subsection{EROS-2 time sampling of the LMC observations}
\label{Sect:timeSampling}

\begin{figure}
	\includegraphics[width=\columnwidth, height=7cm]{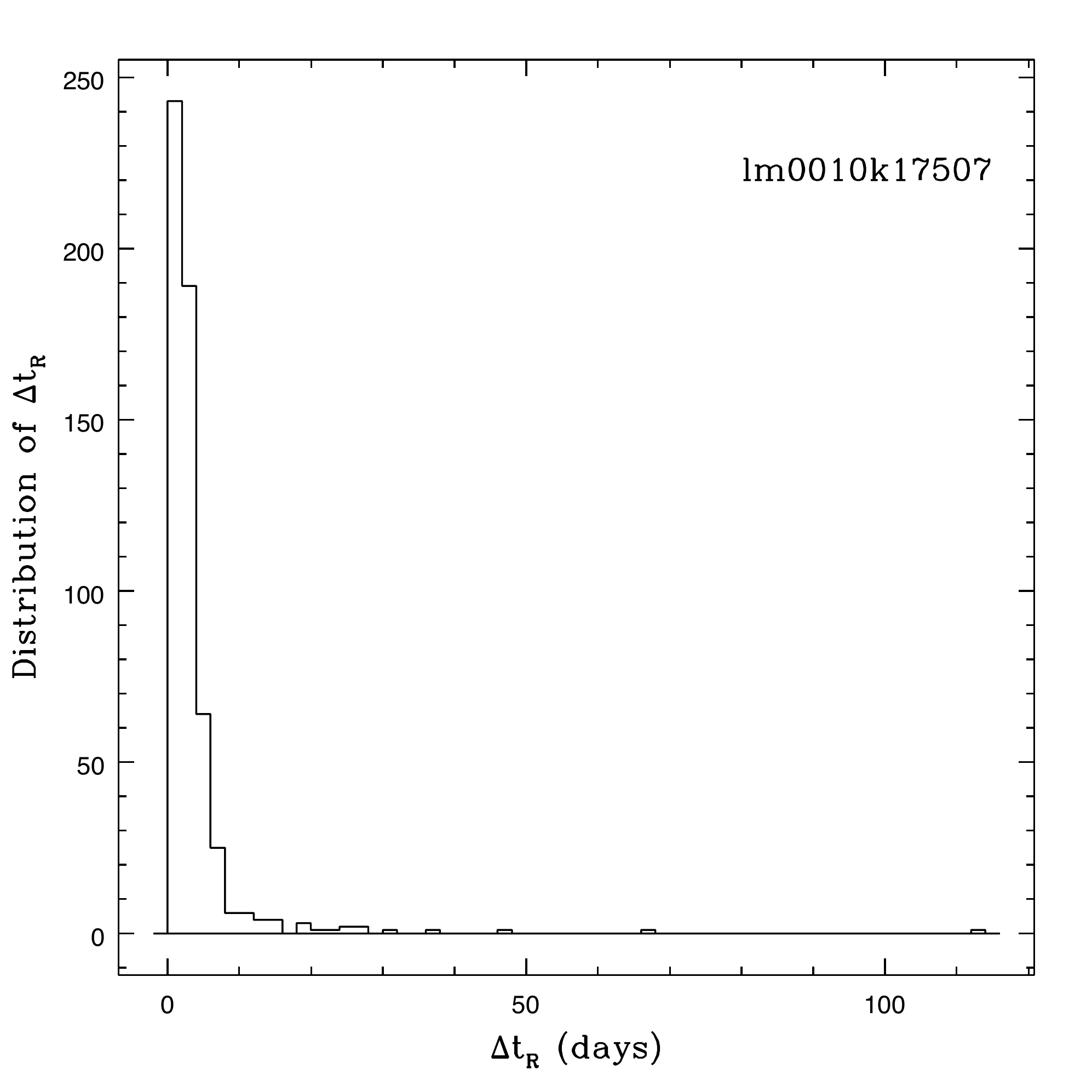}
   	\caption{Distribution of the time interval between two successive observations of the star lm0010k17507, in $R_E$ band, for which a total of 556 good observations were recorded over 2061 days.}
   	\label{deltat}
\end{figure}

The 88 fields constituting the EROS-2 LMC database were rather homogeneously observed during the visibility periods of the LMC throughout the survey.
A typical distribution of the time interval between two successive observations is illustrated in Fig.~\ref{deltat} for the star lm0010k17507, whose full light curve is shown in Fig.~\ref{Fig:abbeLmcLightcurvef} (star nomenclature follows the rules defined by \citet{Derueetal02}).
The majority of observations are separated by two days and most of them by less than five days.
The few long time intervals between 30 and 112 days represent gaps of from weeks to months in the data caused by the LMC not being observable or without good enough air mass conditions from the EROS-2 site.
This distribution of time intervals ensure that the EROS-2 survey is suitable for the detection of all kinds of known LPV stars, with typical periods ranging from 20 to 2\,300 days according to the GCVS \citep{Samusetal09}.
%----------------------------------------------------------------------------------------
\subsection{Data filtering}
\label{Sect:dataFiltering}
In principle, all data in the EROS-2 time series with a magnitude lower than 99.999 and an error value smaller than 9.999 should be assumed to be good.  This typically results in about 389 good observations per star over 2\,002 days in the $R_E$ band, and 477 good observations per star over 2\,324 days in the $B_E$ band.

\begin{figure}
	\includegraphics[width=\columnwidth]{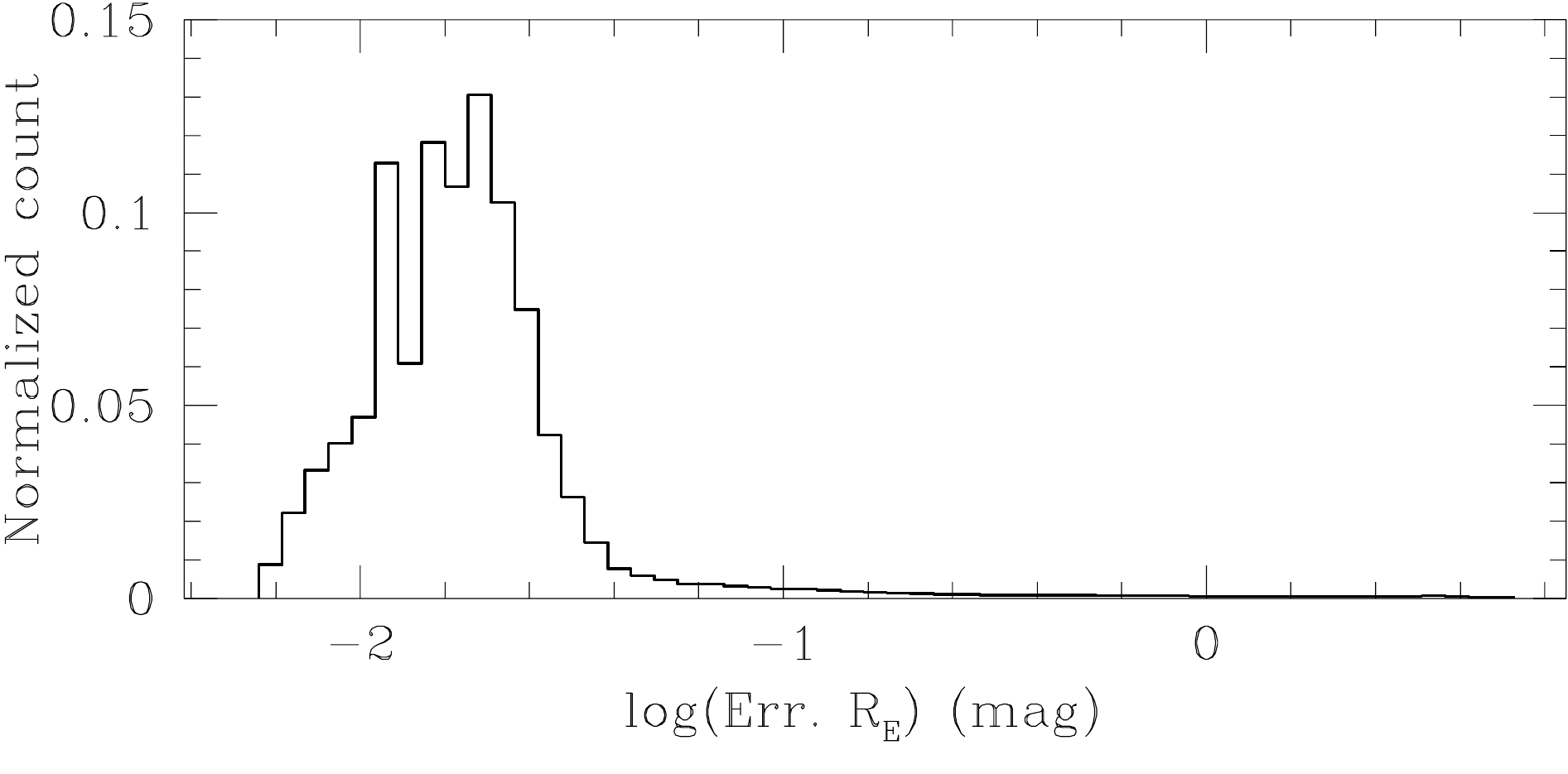}
   	\caption{Normalized distribution of 18\,609\,187 error values, on a log scale, from variables randomly taken, in $R_E$ band.}
   	\label{HistoErrR}
\end{figure}

However, we applied additional filters to the data in the following way:
\begin{itemize}
\item First, we kept only variables with $R_E<$19 mag. These variables are expected to have reliable photometry, as the photometric precision reaches 10\% at $19^{th}$ magnitude \citep[cf][]{ThesisTisserand}. This allowed us to study variables as faint as the magnitude of the red giant clump.
\item Second, we removed all measurements performed during the first 50 days of the mission, since some of these data points have very small error values in the EROS-2 catalog, yet contain some obvious outliers.
\item Third, we eliminate in the remaining measurements the most extreme magnitudes around the mean magnitude by removing one percent of the points from the highest and the lowest values. 
\item Fourth, we eliminated all points with an error larger than one magnitude. 
We show however in the normalized distribution of more than 18 million error values from randomly taken variables (cf. Fig.~\ref{HistoErrR}), that only 0.64\% of the points have an error larger than one magnitude.
Removing points with large errors should not affect our analysis of LPVs, while it can remove outliers that may be present in the data.
\item Finally, we consider only sources with at least 40 points in their filtered $R_E$ light curve.
\end{itemize}

%________________________________________________________________________________________
\section{Identification of LPVs}
\label{Sect:Identification}

In general, it is found that LPVs have photometric variability periods ranging from a few weeks to several years.
Since the typical time interval between two successive EROS-2 observations of the LMC is shorter than five days (see Sect.~\ref{Sect:timeSampling}), the time series of LPVs are expected to vary smoothly with time in the EROS-2 database.
We take advantage of this feature in the first step of our identification of LPV candidates by using a selection criterion based on the Abbe statistic \citep{vonNeumann41} (Sect.~\ref{Sect:Abbe}).
We then exclude in a second step non-LPV candidates from this list by using the color of the candidates (Sect.~\ref{Sect:AbbeLimitForLPVs}).
%---------------------------------------------------------
\subsection{The Abbe test}
\label{Sect:Abbe}

%------------------------------------
\subsubsection{Abbe statistic}
\label{Sect:AbbeStatistic}

\begin{figure}
	\includegraphics[width=\columnwidth, height=8cm]{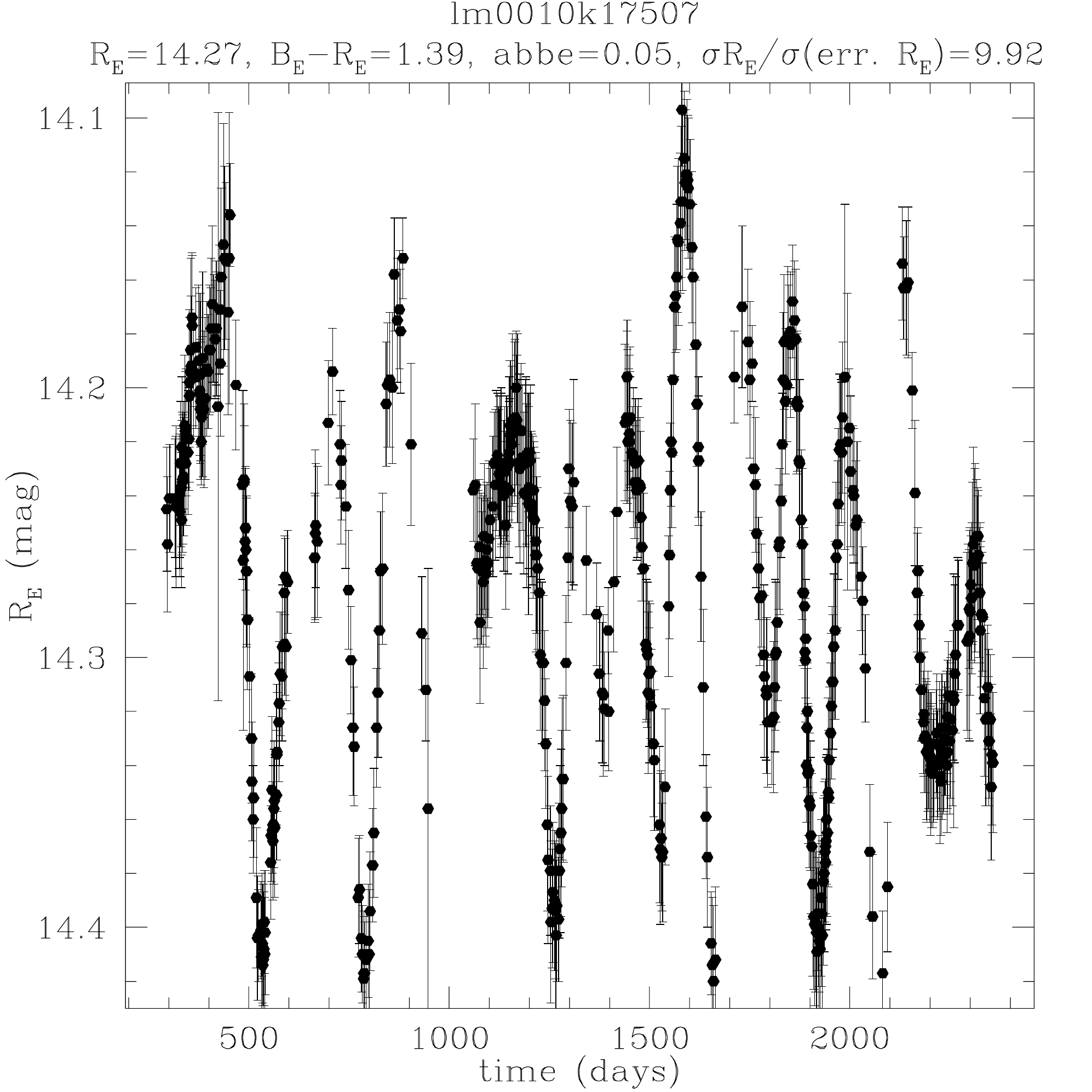}
   	\caption{Time series of the LMC star lm0010k17507 in the $R_E$ magnitude. Times are in Heliocentric Julian day (HJD) - 2\,450\,000.}
   	\label{Fig:abbeLmcLightcurvef}
\end{figure}

\begin{figure}
	\includegraphics[width=\columnwidth]{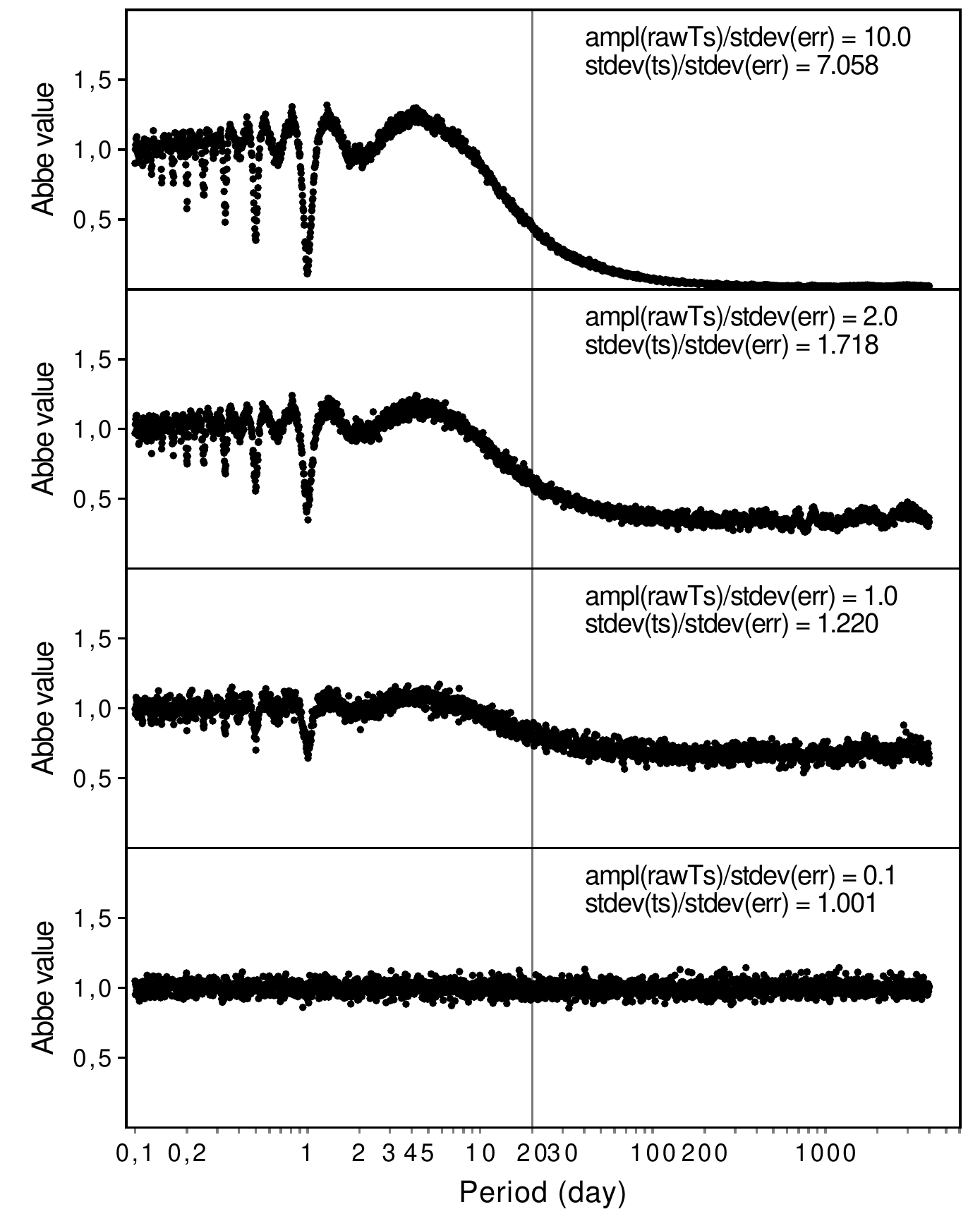}
   	\caption{Values of the Abbe statistic for simulated cosine curves using the time sampling of the star lm0010k17507 displayed in Fig.~\ref{Fig:abbeLmcLightcurvef}. The meaning of ampl(rawTs) and stdev(err) is given in Sect.~\ref{Sect:AbbeIrregularSampling} and that of stdev(ts) in Sect.~\ref{Sect:AbbeLimitForLPVs}. The x-abscissa is the period of the cosine in days, shown on a logarithmic scale. The vertical line indicates the period at 20 days.}
   	\label{Fig:abbeLmcSampling}
\end{figure}

We use the Abbe statistic to identify smoothly varying time series.
We consider a time series $(t_i, x_i)$, where $x_i$ is the value at time $t_i$ of a measured quantity, with index $i$ running over the number of measurements $n$.
In our case, the measured quantity is the $R_E$ or $B_E$ magnitude.
The Abbe statistic considers the differences $(x_{i+1}-x_i)$ between two successive measurements, computes the sum of the square of those differences, and normalizes the result with the standard deviation of the time series and the number of observations.
The Abbe statistic, which we hereafter simply call the Abbe value, is defined to be

\begin{equation}
  r= \frac{n}{2(n-1)} \frac{\sum_{i=1}^{n-1}(x_{i+1}-x_i)^2}{\sum_{i=1}^{n}(x_i-\bar{x})^2}\,,
\label{Eq:Abbe}
\end{equation}
where $\bar{x}$ is the mean value of the measured quantity.
The smoother the time series, the smaller the sum of the differences for a given standard deviation and number of observations.

\subsubsection{Abbe statistic of irregularly sampled time series}
\label{Sect:AbbeIrregularSampling}
The value of Eq.~(\ref{Eq:Abbe}) for a time series depends on the variability timescale of the light curve and the time sampling.
In real data, the time sampling is not uniform.
To assess the impact of the irregular time sampling on the Abbe statistic, simulations of time series of cosine functions $x_i = \cos(2\,\pi\, t_i/P)$ were performed, with a time sampling typical of the EROS-2 observations of the LMC.
We take for this purpose the time sampling of the star lm0010k17507, whose light curve is shown in Fig.~\ref{Fig:abbeLmcLightcurvef}.

Results are shown in Fig.~\ref{Fig:abbeLmcSampling}. The simulations were performed with periods ranging from 0.1 to 4000 days. Each point in the figure represents the Abbe value for one periodic time series, where the period is plotted on the x-axis.
From the top to bottom panel, an increasing level of noise is added to the time series. The error added to each point in the time series is computed from a gaussian distribution, of which the standard deviation $\sigma_\mathrm{err}$ (noted as stdev(err)) relative to the peak-to-peak variability amplitude $A_\mathrm{var}$ of the error-free cosine values of the simulations (noted as ampl(rawTs)) is reported in each panel of the figure.
We define the noise level as $A_\mathrm{var} /\sigma_\mathrm{err}=ampl(rawTs)/stdev(err)$.
When the noise level increases, the amplitude of the computed Abbe values becomes smaller. For the noisiest simulations, i.e. with $A_\mathrm{var} /\sigma_\mathrm{err}=0.1$, the time series are so dominated by noise that no correlation remains between the values of two successive points in the time series.
In that case, the sum of the square differences $(x_{i+1}-x_i)^2$ between two successive points equals, on average, twice the standard deviation, leading to a value of $r \simeq 1.0$.
This conclusion is supported by the work of \citet{Strunov06}, who shows that a normal distribution of uncorrelated values in a time series with more than 60 points leads to Abbe values that are normally distributed with a mean at 1.0 and with a variance of $\sqrt{\frac{n-2}{(n-1)(n+2)}}$.

The simulations with EROS-2 time sampling also demonstrate that the amplitudes of the oscillations in the Abbe values decrease towards shorter periods, which can be seen as a decrease in dispersion around unity, and the Abbe values decrease towards zero for longer periods.
This is important when selecting LPV candidates characterized by long periods.

\begin{figure}
	\includegraphics[width=\columnwidth,height=5.5cm]{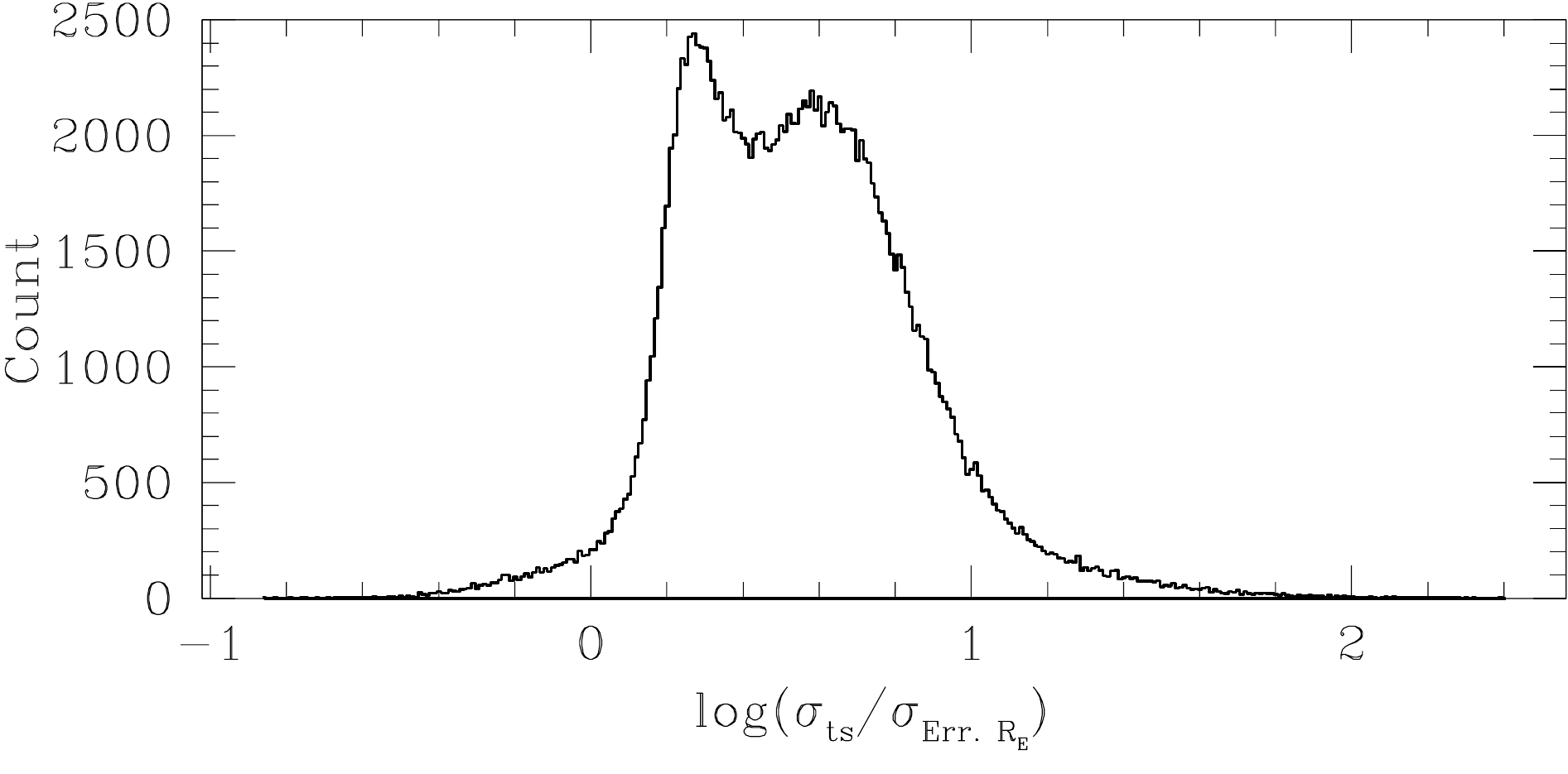}
   	\caption{Distribution of the ratio $\sigma_{ts}/\sigma_{Err. R_E}$ on a log scale, for a sample 20\% of which are LMC EROS-2 variables that are randomly selected (172\,427 stars).}
   	\label{StatNoiseSample}
\end{figure}

%------------------------------------
\subsubsection{Abbe statistic and long period variables}
\label{Sect:AbbeLimitForLPVs}

\begin{figure}
	\includegraphics[width=\columnwidth,height=6cm]{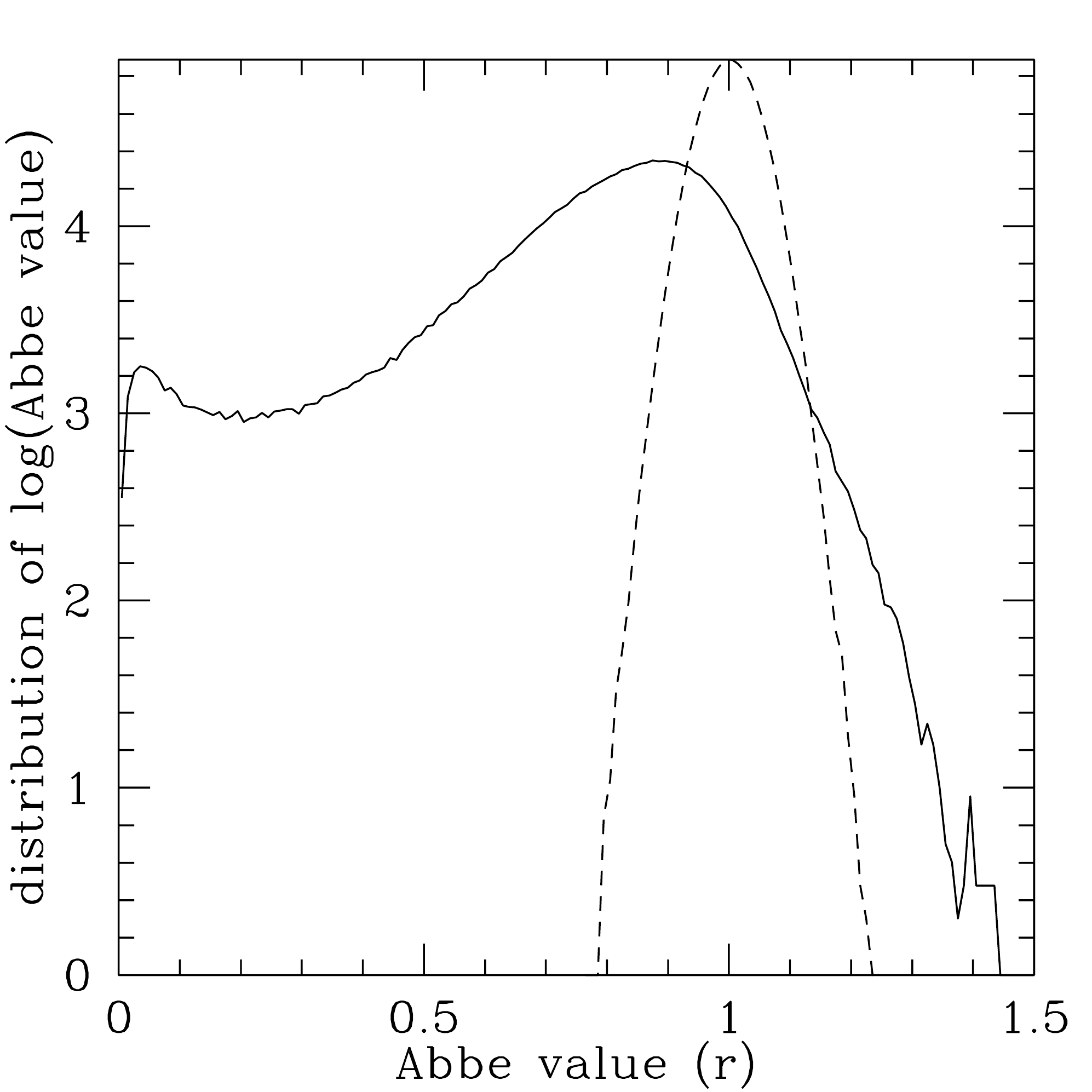}
   	\caption{Distribution of the Abbe statistic for a cleaned time series of all LMC variable stars (solid line).
	         The dashed line shows the expected distribution of the Abbe values for simulated time series with normally distributed magnitudes.}
   	\label{DistAbbe}
\end{figure}

From the conclusion of the section above, we can infer on a limit to the Abbe value $r_\mathrm{lim}$ below which mainly variable stars with long variability timescales are selected.
Fig.~\ref{Fig:abbeLmcSampling} shows that if $r_\mathrm{lim}$ is properly chosen, only a small fraction of the candidates with $r< r_\mathrm{lim}$ will have short periods, of either around one day or, to a lesser extend, around 0.5~days (both due to the one day aliasing introduced in the time series obtained from a ground-based survey).

 The Abbe value limit depends on the $A_\mathrm{var} /\sigma_\mathrm{err}$ level of the time series.
Unfortunately, this quantity is unknown from observations, since we do not know the variability amplitude $A_\mathrm{var}$ of the noise-free light curve.
Therefore, instead of using $A_\mathrm{var} /\sigma_\mathrm{err}$, we use the ratio $\sigma_\mathrm{ts}/\sigma_\mathrm{err}$ where $\sigma_\mathrm{ts}$ is the measurable standard deviation of the observed time series (noted as stdev(ts) in Fig.~\ref{Fig:abbeLmcSampling}).
The mean value of this ratio over all simulated time series is reported in each panel of Fig.~\ref{Fig:abbeLmcSampling}, and equals 7.1, 1.7, 1.2, and 1.0 for $A_\mathrm{var} /\sigma_\mathrm{err}=10$, 2, 1, and 0.1, respectively. Fig.~\ref{StatNoiseSample} shows the distribution of the ratio $\sigma_\mathrm{ts}/\sigma_\mathrm{err}$ for a sample of variables. The mean value of the distribution is equal to 5.89, hence our signal-to-noise ratios from the simulations are compatible with the values computed from the real data.
The value of $\sigma_\mathrm{ts}/\sigma_\mathrm{err}$ for the star lm0010k17507 is 9.92. If we consider the time sampling of this star to be representative of LMC LPVs in the EROS-2 database, we expect all periodic stars with periods longer than 20~days to have $r<0.4$ (see top panel of Fig.~\ref{Fig:abbeLmcSampling}). 
The population of stars satisfying this criterion would be only slightly polluted by periodic stars with periods of one day or its sub-multiples, as discussed previously.

The distribution of Abbe values for all time series in the EROS-2 database of variable stars is shown as a solid line in Fig.~\ref{DistAbbe}.
Two populations are distinguishable in this distribution.
The first one which peaks at $r \simeq 0.04$, should correspond to the long timescale variable stars to which LPVs belong.
The second population, around $r=1.0$ corresponds to the time series that do not display a smooth variation in magnitude from one observation to the next. 
The condition $r<0.4$ suggested by the analysis of the Abbe statistic in Sect.~\ref{Sect:AbbeIrregularSampling} is seen in Fig.~\ref{DistAbbe} to help us distinguish between the two populations to include as much of the first population as possible.

\begin{figure}
	\includegraphics[width=\columnwidth]{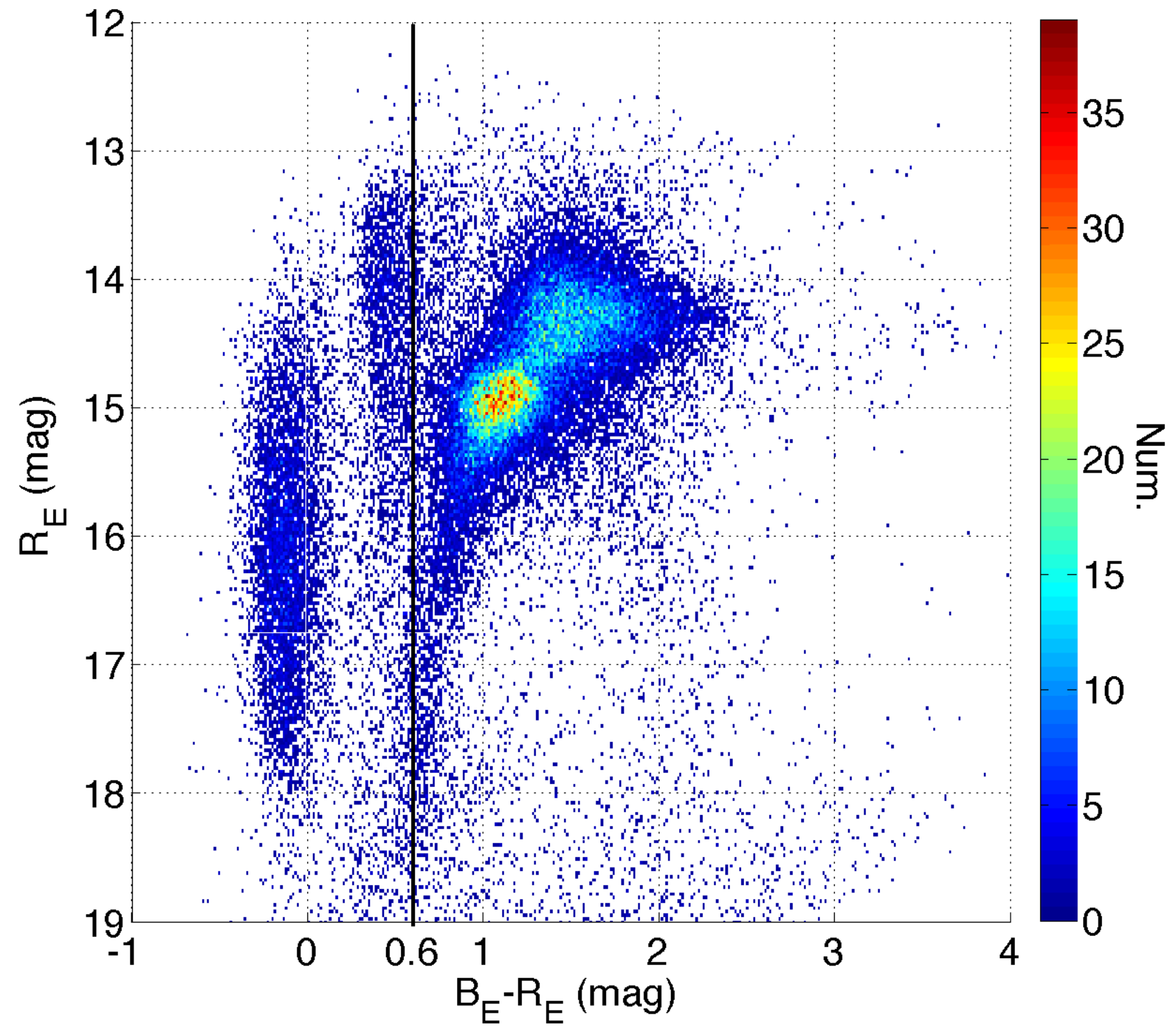}
   	\caption{Number density in the EROS-2 LMC CMD for variable stars with an Abbe value $<$0.4. The vertical line at $B_E-R_E=0.6$ mag shows the limit chosen for selection of LPV candidates. The CMD is divided into 250x357 bins of size 0.014x0.028 mag. The number of stars in each bin is plotted according to the color-scale shown on the right of the figure.}
   	\label{CMD_candlpv}
\end{figure}

The condition $r < 0.4$ on the Abbe value leads to the selection of 60\,723 candidates from the LMC variable stars, in which LPVs are to be found.
We plot these candidates on the ($B_E-R_E, R_E$) CMD of Fig.~\ref{CMD_candlpv}. We can safely assume that most of the red-giant variable stars matching our long-period variability criteria were selected. On top of that, bluer variable types such as Be stars or long-period Cepheids for instance are also selected. We use the color of the red giant clump at $B_E-R_E=0.6$ mag as the limit of the RGB, hence variable stars with a $B_E-R_E>0.6$ mag were chosen as potential LPV stars. The sample created consists of 43\,583 EROS-2 sources.
The criteria $r >0.4$ and $B_E-R_E>0.6$ mag selects 280\,769 variable stars, indicating that variable stars redder than the red clump are not necessarily LPVs according to our criteria. A look at the hundred brightest light curves shows that they include: (1) binary stars, (2) stars with small amplitude and short variability, and (3) variables that have a long-term variability component associated with a secondary variability component that is too short to result in a light curve smooth enough to give a small Abbe value.

\section{Period search}
\label{Sect:Periodsearch}
\subsection{Methods}
\label{SubSect:Methods}
 
\begin{figure*}
	\centering
	\includegraphics[width=14cm]{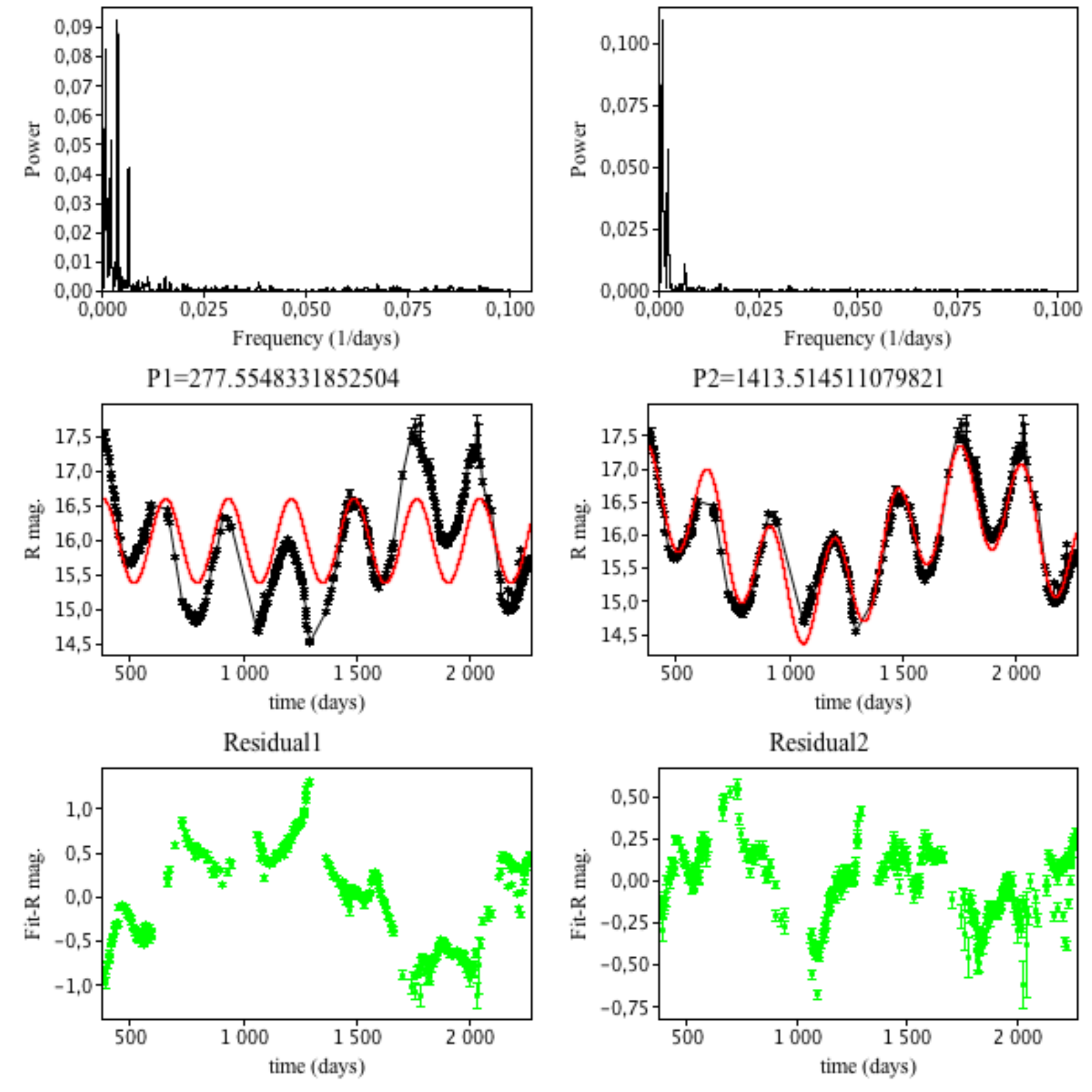}
   	\caption{Period search example for star lm0024n4304. The top panels show the successive power spectra of the raw light curve (left panel) and of the first  residual (right panel). The middle panels display the light curve in R band (black points with error bars), with the fitted Fourier series superimposed (red curve) for the first (left panel) and first and second (right panel) periods found. The bottom panels show the residuals after subtraction of the fits from the light curve.}
	\label{example}
\end{figure*}

We search for periods characterizing the light curves of our LPV candidates using Deeming \citep{Deeming75}, Lomb-Scargle \citep{Press07}, and non-weighted generalized Lomb-Scargle \citep{Zechmeister09} methods.
The Deeming method uses the Fourier transform applied to unevenly sampled light curves. The Lomb-Scargle method is derived from the least squares fitting of sine and cosine functions with a fixed mean value. Both methods are quite similar and should provide similar results, hence they are used as a double-check. The non-weighted generalized Lomb-Scargle method adds a floating mean value computed from a sine function fitted to the time series as an extra parameter to the classical Lomb-Scargle least squares. It prevents the selection of an incorrect period due to a poor phase coverage.
 For each of these period search methods, we use an iterative procedure to extract the significant periods that characterize a light curve.
The standard deviation $\sigma$ in the periodogram provides an estimate of the noise level.
The significance of each peak in the periodogram is then evaluated relative to this standard deviation $\sigma$. A peak is considered significant if its value is at least 3 $\sigma$ from the mean value of the periodogram.
We follow a classical period-search procedure, i.e we first search for the most significant frequency in the periodogram of the original light curve (Fourier transform in the case of Deeming for example), fit the light curve with a sine function, and compute the residual from this fit.
We then proceed similarly with the residual time series so obtained, and repeat the process until either no additional significant period is found or a maximum of twelve periods are found.
Figure~\ref{example} illustrates these steps with an example for the star lm0024n4304 in the R band. We show in the top panels periodograms resulting from the period search, the resulting model of the fit upon which we superimpose the raw light curves in the middle panels, and in the bottom panels the corresponding residual light curves after subtracting the fit from the raw light curve according to the period(s) found. 
We note that the periodic light curves that are not sinusoidal will have their harmonics extracted during this process, so that the different periods may not be independent of each other.
In our approach, those harmonics will be considered as separate periods.

Numerically, the periodograms are computed for frequencies ranging from 1.818.10$^{-4}$ to 0.1 day$^{-1}$, corresponding to periods from 10 to 5\,500 days (2.5 times the mean EROS-2 duration) with frequency steps computed as  \begin{equation}
	$d$\nu(t)$=$\frac{1}{\Delta T.N}$$,
\end{equation}
where $\Delta T$ is the time span of the time series and $N$ is an oversampling factor that we set to 300. As most EROS-2 time series on the LMC have a duration around 2\,000 days, we usually have a d$\nu(t)$$\simeq2.10$$^{-6}$day$^{-1}$.

\begin{figure}
	\includegraphics[width=\columnwidth]{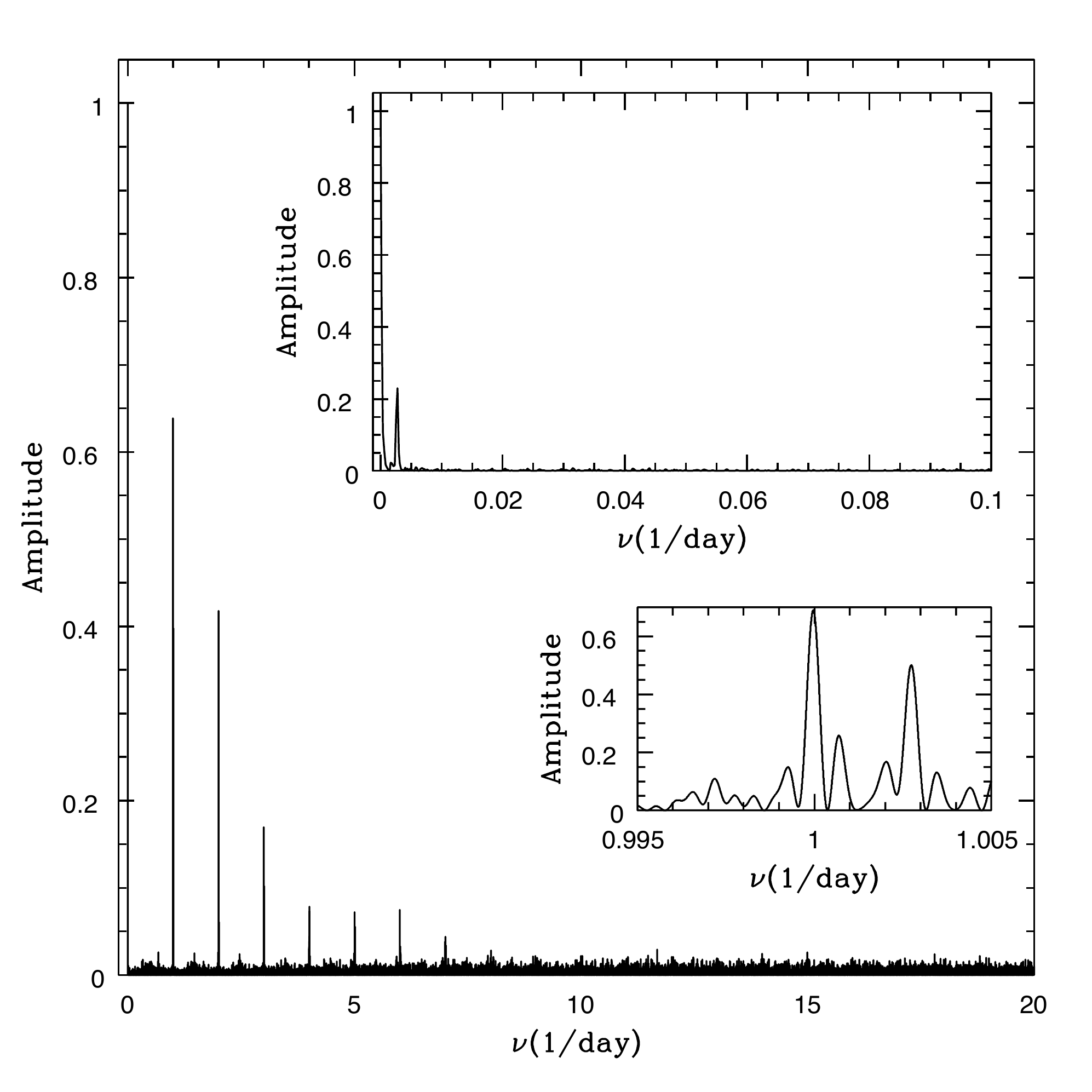}
   	\caption{Spectral window of the star lm0396l14700. Top insert: zoom into the frequency interval used for period search. Bottom insert: Structure of the spectral window close to the one day alias.}
   	\label{SW}
\end{figure}

	In Fig.~\ref{SW}, we show a typical spectral window for the LMC EROS-2 sources. Apart from the peak at the origin, the largest amplitude peaks are at frequencies of one day$^{-1}$ and higher. In the case of the period search for short-period variables (of period shorter than one day) these peaks will produce aliases in the power spectrum. In the bottom insert, we show the spectral window around one day$^{-1}$. We can see that the sampling allows us to clearly define the structure of the peaks. The two highest ones are at one solar day (1 day$^{-1}$), and one sidereal day (1.0027 day$^{-1}$). As we applied a selection criterion to the Abbe values, only LPVs were retained. Spectral window for the frequency interval used for the period search for our LPV candidates is shown in the top insert of Fig.~\ref{SW}. The peak clearly visible at a frequency of 0.0027 day$^{-1}$ is produced by the one-year artifact. Its normalized amplitude is smaller than 0.25. The aliases created in the periodogram then, are not expected to be significant. According to the values of the peaks found at frequencies lower than 0.1 day$^{-1}$, we can assume that the period(s) found for our LPV candidates are not produced by aliases.

\subsection{Results}
\label{Sect:Results}

\begin{figure}
	\includegraphics[width=\linewidth, height=5cm]{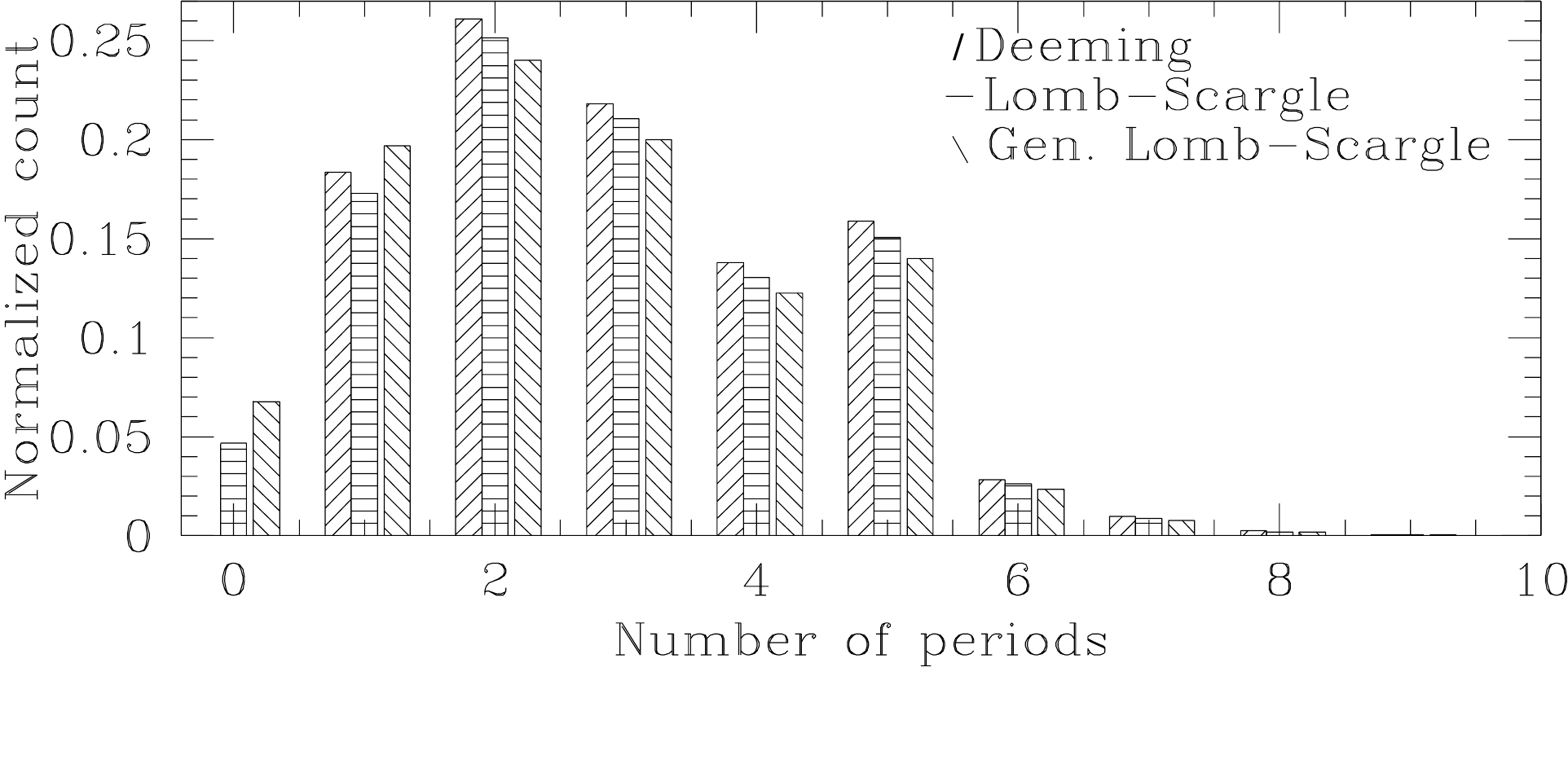}
   	\caption{Normalized histogram of the number of periods found for the Deeming, Lomb-Scargle and generalized Lomb-Scargle methods.}
   	\label{HistoNPerMethod}
\end{figure}

\begin{figure}
	\includegraphics[width=\linewidth, height=4.5cm]{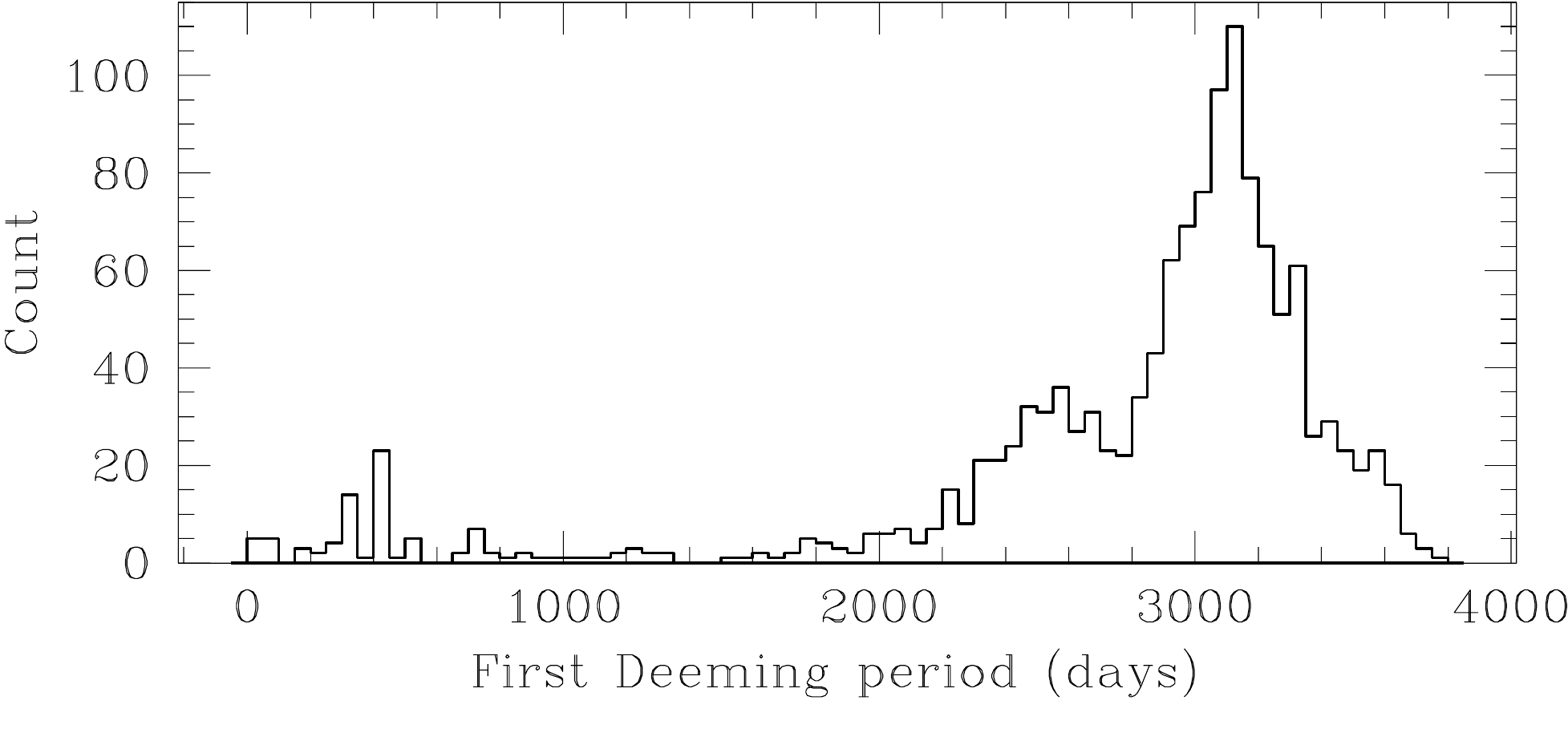}
   	\caption{Histogram of the first period found with the Deeming method, for variables for which a period was found with neither the Lomb-Scargle nor generalized Lomb-Scargle methods.}
   	\label{HistoP1Deem_NoPLSGLS}
\end{figure}

\begin{figure}
	\includegraphics[width=\linewidth]{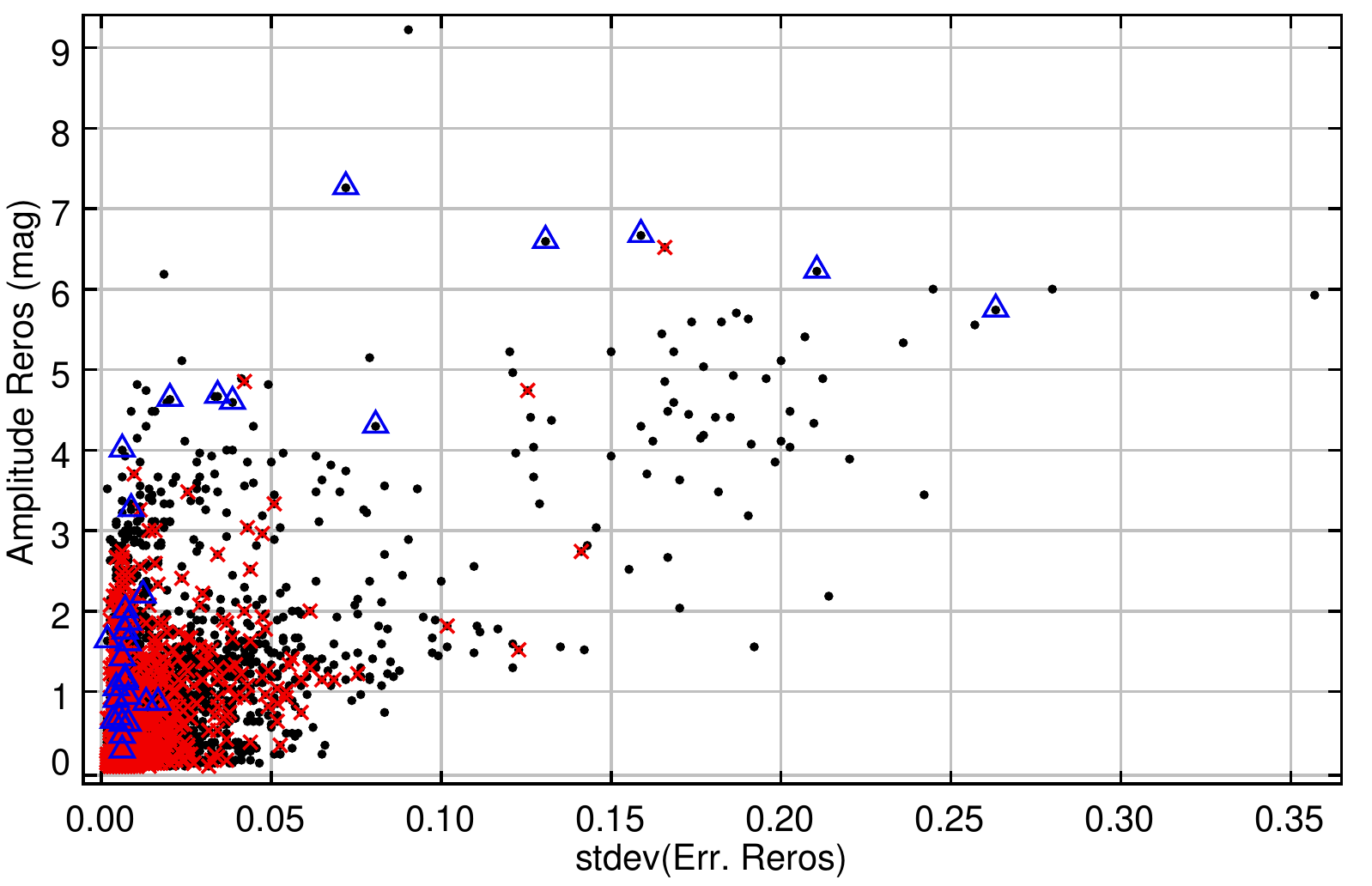}
   	\caption{Peak-to-peak amplitude versus the standard deviation in the error values. Stars without any period simultaneously found with the three-period search methods, are shown as black points. Red crosses are stars showed in Fig.~\ref{HistoP1Deem_NoPLSGLS}. Blue triangles are RCB and DY PER identified by  \citet{Tisserandetal09}.}
   	\label{Sigerr_AmpR_NoPerLSGLS}
\end{figure}

Figure \ref{HistoNPerMethod} shows a normalized histogram of the number of periods found for each of the three methods, according to the criteria discussed in the previous section. Each star appears only once, which means for example that a star for which three periods are found is not counted in the two-period bin. The proportions of variable stars with the same number of periods appear to be comparable independently of the method used. However, for roughly six percent of stars no period was found with the Lomb-Scargle and the generalized Lomb-Scargle methods, whereas Deeming found at least one.
We select these 1\,244 variables stars and show the histogram of their first period found with the Deeming method in Fig.~\ref{HistoP1Deem_NoPLSGLS}. We can see that for the majority of these stars the period exceeds 2\,000 days, which is the mean EROS-2 time length. Thus, a complete phase coverage is impossible in these cases. Nevertheless, visual inspection of these stars shows that most of them exhibit clear long-term variability features.\\

For each period-search method, we consider up to five periods. If three period values, each found by a different method, are within a range of $\pm$10\% we define a new period value that is equal to the mean of the three corresponding periods determined by the different methods.

For almost 14\% of our LPV candidates, no common period was found by all three methods. Visual inspection of these time series shows that for about 10\% of them we find a typical LPV light curve. A cause of their exclusion from the sample is that the highest peak in their periodogram is at an harmonic period for one of the methods, while in other cases it is at the fundamental period ($3\%$ of the cases). We reclassify these stars by choosing the period of the fundamental mode. Figure ~\ref{Sigerr_AmpR_NoPerLSGLS} shows the peak-to-peak amplitude in the $R_E$ band versus the standard deviation in the error values for the remaining $90\%$ of the stars where no clear LPV feature was observed. Most of them appear to have a small standard deviation in the error values ($\sigma_{err}<0.05$), even if a small group of variables is visible at large standard deviation with a larger mean amplitude ($\sigma_{err}>0.1$ and $R_E$ Amplitude$>1.5$ mag). A visual inspection of the 72 variables with $\sigma_{err}>0.1$ reveals that 47 of these stars are affected by data reduction problems. As a consequence, many outliers are visible in these light curves, which explains the higher amplitude values. These outliers have errors smaller than expected for their magnitude, which cause the larger standard deviation value in the errors. The resulting variability is then very unreliable. We therefore exclude them from further study.

 Visual inspection reveals that some non-periodic variables are present in our sample. We identified 32 variables of R Coronae Borealis (RCB) and DY Per types already known in EROS-2 and listed in \citet{Tisserandetal09}. We show them in Fig.~\ref{Sigerr_AmpR_NoPerLSGLS} as triangles. These stars are flagged in the online catalog containing 43\,583 EROS-2 sources (see Sect.~\ref{Sect:catalog}) but are not taken into account for the following study.
In Fig.~\ref{Sigerr_AmpR_NoPerLSGLS}, we also show, as red crosses, the 1\,244 previously identified variables for which only the Deeming method find period(s). These stars lie in the group with the smaller error dispersion, which lends support to the assumption that variables with small $\sigma$$_{err}$ are probably genuine variables with a very long period.\\
We then obtained 6\,008 variables without finding any period and 37\,543 variables (86\% of the LPV candidates) for which at least one period value is retrieved, distributed as follows: 9\,666 variables with a single period retrieved, 9\,904 with two periods, 8\,595 with three periods, 5\,845 with four periods, and 3\,533 with five periods. Finding LPVs with such a large number of periods is unexpected, but is probably caused by our keeping harmonics as independent periods.
We created a catalog with 43\,551 LPV candidates (see Sect.~\ref{Sect:catalog}). The sample of LPVs studied in the following sections is composed of the 37\,543 variable stars for which at least one period was found.

\begin{table}
\caption{Results of cross-identifications between EROS-2, OGLE III, and MACHO LPV catalogs.}             % title of Table
\label{tableCM}      % is used to refer this table in the text
\centering                          % used for centering table
\resizebox{\columnwidth }{!}{ 
\begin{tabular}{r r r r r}
\hline
Catalog & Number of & Common LPVs & LPVs found & LPVs found\\    
  ~    & LPVs in cat.& (EROS-2, cat.) & in EROS-2 only & in cat. only\\
\hline
OGLE III & 91\,995 &  27\,940 & 15\,643 & 64\,054\\
MACHO & 56\,453 & 25\,812 & 17\,771 & 30\,641\\
\hline
\end{tabular}}
\end{table}

We performed cross-identifications of all our LPV candidates with two LPV catalogs available from both the OGLE III \citep{Soszynskietal09} and MACHO \citep{Fraseretal08} surveys. As some of the EROS-2 RCB and DY Per stars are considered as LPVs in both the OGLE-III and MACHO catalogs, we kept them for these cross-identifications. A two arcsec search radius was used to identify counterparts to our LPV candidates in these two catalogs. For cases where one LPV candidate gives two matches, we keep the one with the closest coordinates. The results of this cross-identification are summarized in Table~\ref{tableCM}.

We derived 64\% of our LPVs from the OGLE III catalog. Most of the remaining 36\% of LPVs that were found only in EROS-2 are in fields of the LMC not covered by OGLE III. The large number of LPVs found only in OGLE III correspond mainly to OGLE small-amplitude red-giant variables (OSARGs) \citep{Wrayetal04}, which constitute 86\% of the OGLE III LPV catalog. As shown in \citet{Soszynskietal04b} (see their Fig. 2), OSARGs can have amplitudes as small as few millimagnitudes in I band. Since Fig.~\ref{HistoErrR} shows that most of EROS-2 errors have values between 0.01 and 0.025 mag in $R_E$ band, an identification of all OSARGs as variable stars in EROS-2 is not very likely.

We found that 59\% of the LPVs are in common between EROS-2 and MACHO, this smaller percentage compared to OGLE III being caused by a smaller sky coverage for the MACHO survey. Some of the 30\,641 LPVs found only in MACHO may come from small amplitude LPVs, since 6\,031 stars have an amplitude in R band smaller than 0.025 mag. Another part would consists of the 11\,215 LPVs from the MACHO catalog affected by the one-year artifact (cf. \citet{Fraseretal08}) that are not selected with EROS-2 (See Sect.~\ref{SubSect:CharMono}.). The sky coverage is an important factor as well. Most of the LPVs found only in EROS-2 are in fields of the LMC not covered by MACHO, mainly external fields but also those in the LMC bar.

\subsection{Characterization of LPVs}
\label{Sect:characterization}

\subsubsection{Mono-periodic LPVs}
\label{SubSect:CharMono}

\begin{figure}
	\includegraphics[width=\columnwidth, height=6.6cm]{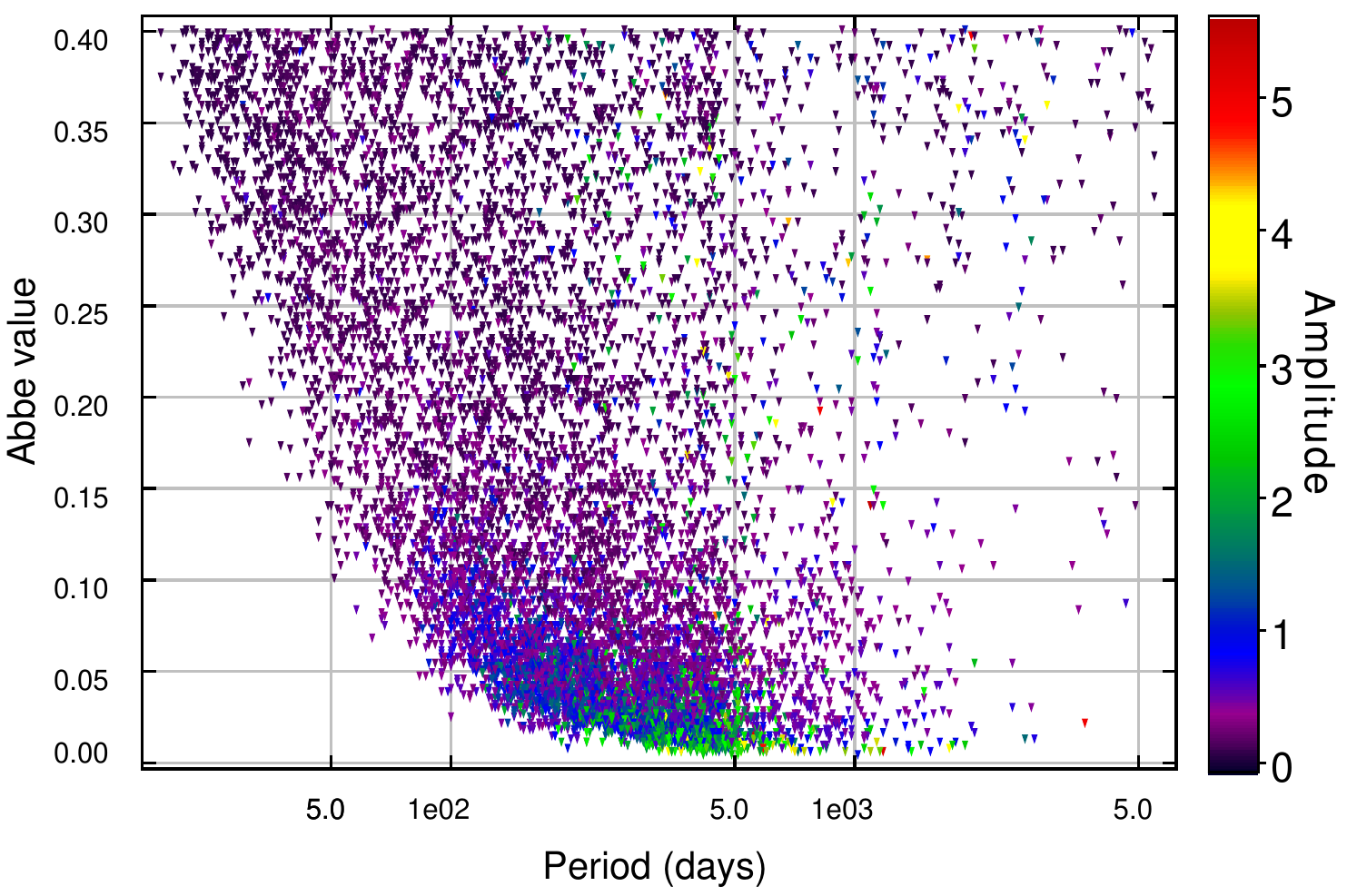}

	\includegraphics[width=\columnwidth, height=6.6cm]{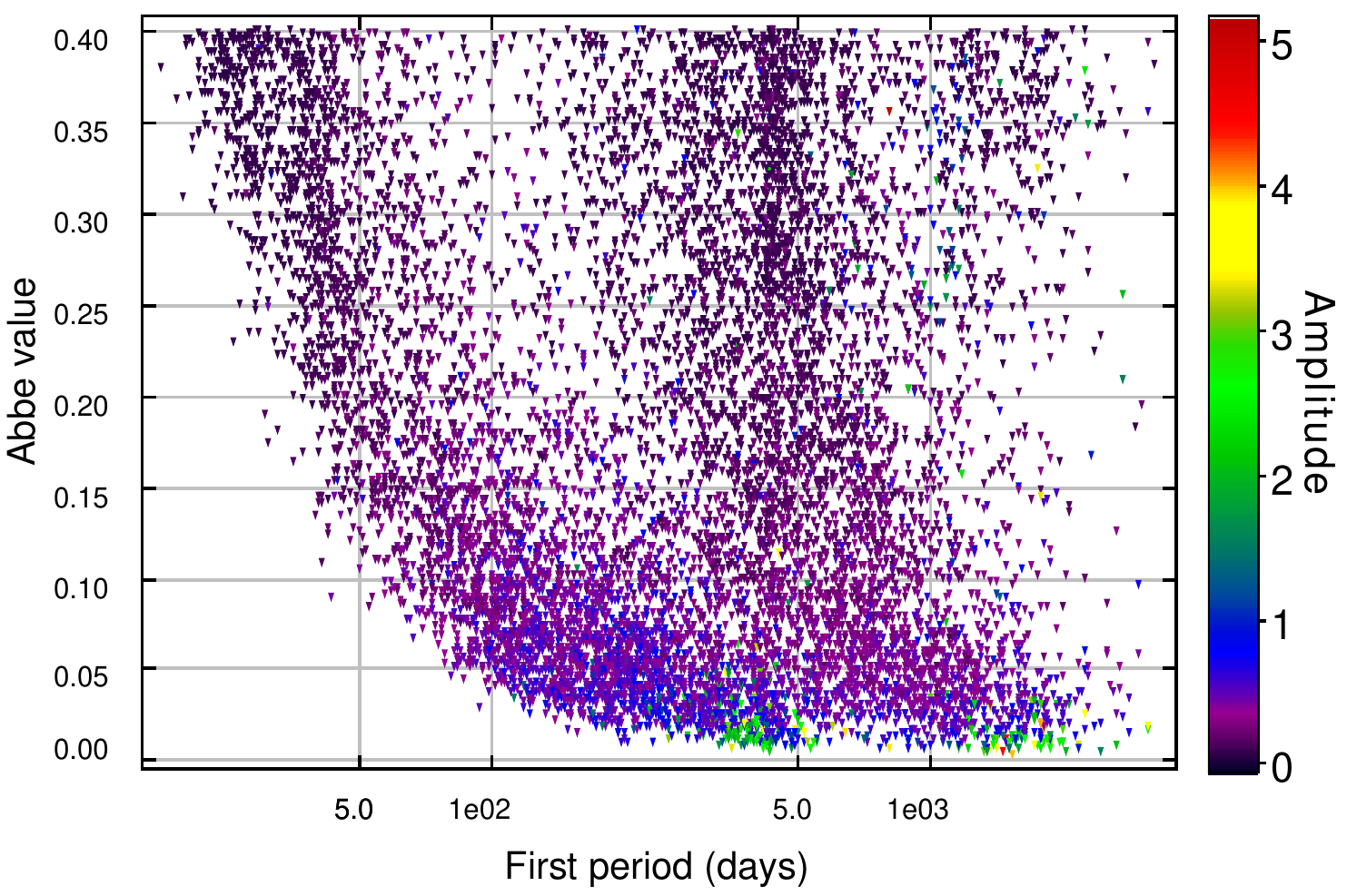}

	\includegraphics[width=\columnwidth, height=6.6cm]{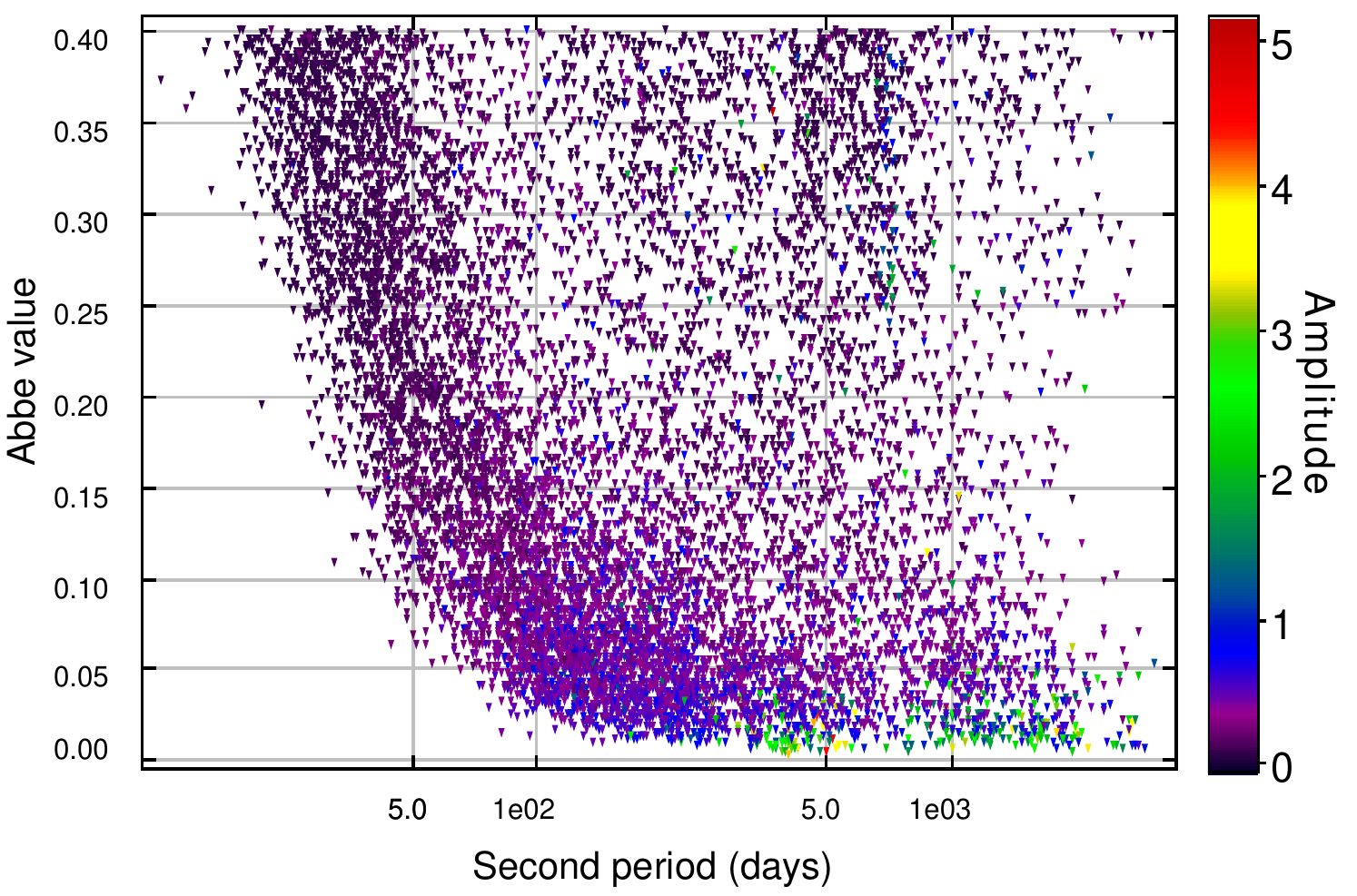}

   	\caption{Top panel: Abbe values versus period on a log scale, for mono-periodic LPV, with color-coded amplitude.
	Middle panel: Abbe values versus first period on a log scale, for double-periodic LPV, with color-coded amplitude.
   	Bottom panel: Abbe values versus second period on a log scale, for double-periodic LPV, with color-coded amplitude.}

   	\label{AbbePerAmp}
\end{figure}

\begin{figure}
	\includegraphics[width=\columnwidth, height=4.5cm]{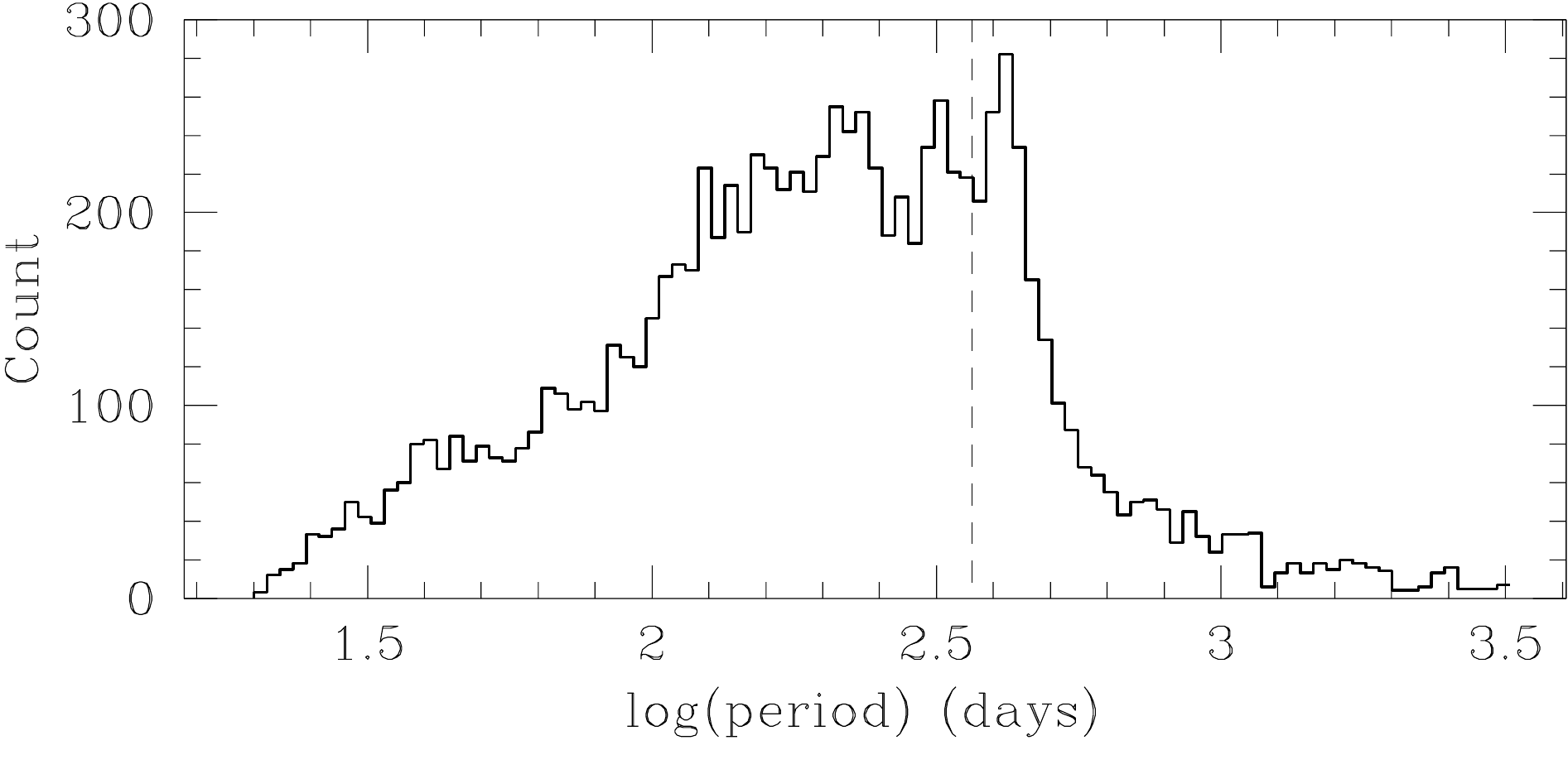}
	
	\includegraphics[width=\columnwidth, height=4.5cm]{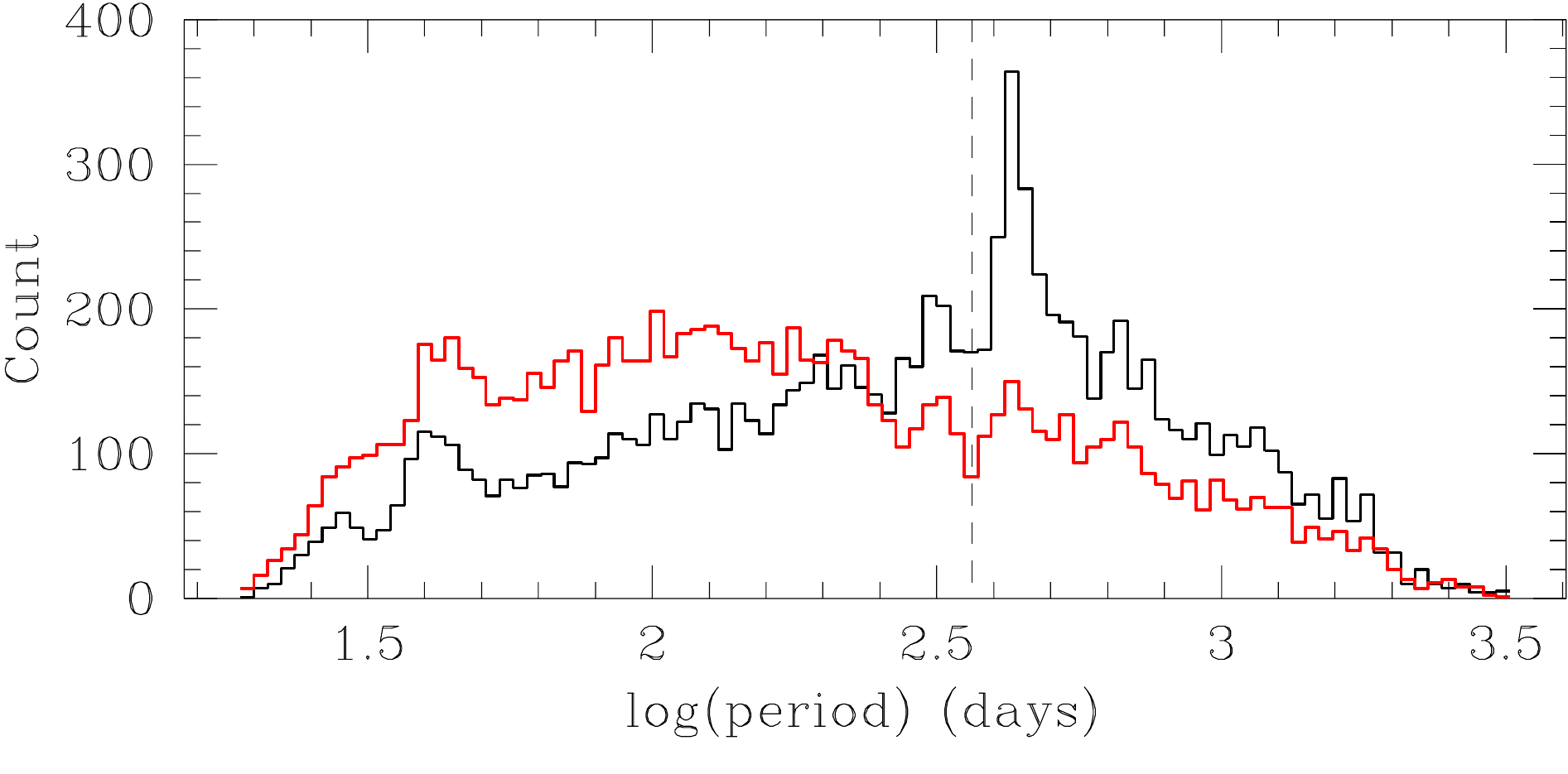}
   	\caption{Top panel: Histogram of the periods for mono-periodic LPVs, on a log scale.
	Bottom panel: Histogram of the periods for double-periodic LPVs, on a log scale. The black line is for the first periods found, the red one is for the second periods found. Vertical dashed lines at log $P=2.56$ indicate the period at one year.}
   	\label{HistoPerDoublePer}
\end{figure}

More than 22\% of our LPV candidates are found to have only one significant period according to the criteria mentioned in the two previous sections (i.e a period with a significant peak value in the periodogram and found to be in agreement with the three period search methods). The Abbe values are plotted against the period with color-coded amplitude in Fig.~\ref{AbbePerAmp}. 
We first note that no variables have a small Abbe value (less than 0.1) and short period (shorter than 50 days), as expected from Sect.~\ref{Sect:AbbeIrregularSampling}.

The distribution of periods in the top panel of Fig.~\ref{HistoPerDoublePer} shows an increase in the number of mono-periodic LPVs up to log $P=2.3$ (P=199.5 days). Then, instead of a monotonic decrease of the distribution towards higher periods, two peaks are visible at log $P=2.51$ and log $P=2.6$ (P=323.6 days and P=398.1 days). Raw EROS-2 time series contain a one year artifact caused by the seasonal effect, which is visible as linear trends in magnitude. In the light curves provided, this artifact was corrected to some extent \citep[cf][]{ThesisTisserand} and explains the smaller number of stars with a period of around one year, as indicated by the vertical dashed line in Fig.~\ref{HistoPerDoublePer}. Variables with periods a few weeks shorter or longer than this one year artifact could still contain these trends in their light curves, hence produce an artificially larger number of LPV candidates with periods around one year.

\subsubsection{Double-periodic LPVs}
\label{SubSect:CharDouble}

For double-periodic variables, the middle and bottom panels of Fig.~\ref{AbbePerAmp} display the same general characteristic as mono-periodic LPVs. The distribution of periods in the bottom panel of Fig.~\ref{HistoPerDoublePer} also shows a larger number of stars close to the one-year artifact, especially for the first periods slightly above one year visible as a strong peak at log $P=2.62$. The distribution of secondary periods is more uniform, mainly for periods between 40 days (log $P=1.6$) and one year.

\section{Main properties of the sample}
\label{Sect:Properties}

As LPVs are known to follow distinctive sequences in the period-luminosity diagram, we identify and discuss here the properties of these sequences for our LPV sample.
The surface abundances of red giant stars can be dominated by either oxygen (O-rich) or carbon (C-rich), depending on their evolutionary phase. Some of them are obscured by circumstellar dust. Visual photometry, then, does not allow a direct comparison of these stars with the lightly obscured ones. Hence, near- and mid-infrared observations are needed to investigate the properties of RGB and AGB stars. Thus, in addition to EROS-2 data, we use 2MASS and \textit{Spitzer} infrared data for the analysis of our candidates.

\subsection{Infrared properties}
\label{IRproperties}

\begin{figure}
	\includegraphics[width=\columnwidth, height=8cm]{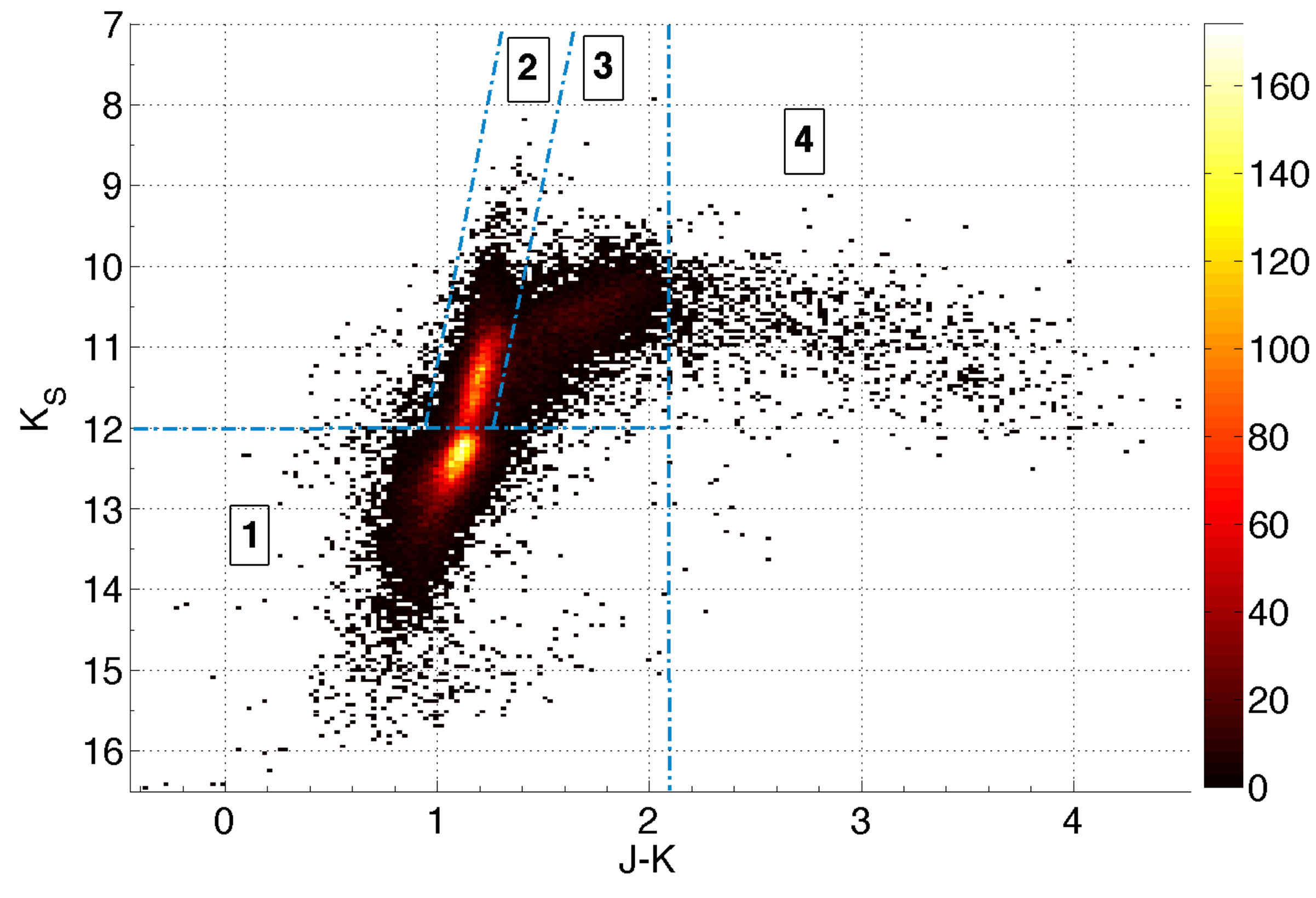}
   	\caption{Number density of the EROS-2 LPV candidates cross-matched with the 2MASS catalog, in the J-$K_{s}$, $K_{s}$ CMD. The CMD is divided into 200x200 bins with a bin size of 0.061x0.043 mag.}
   	\label{KJK_density}
\end{figure}

\begin{figure}
	\includegraphics[width=\columnwidth, height=6cm]{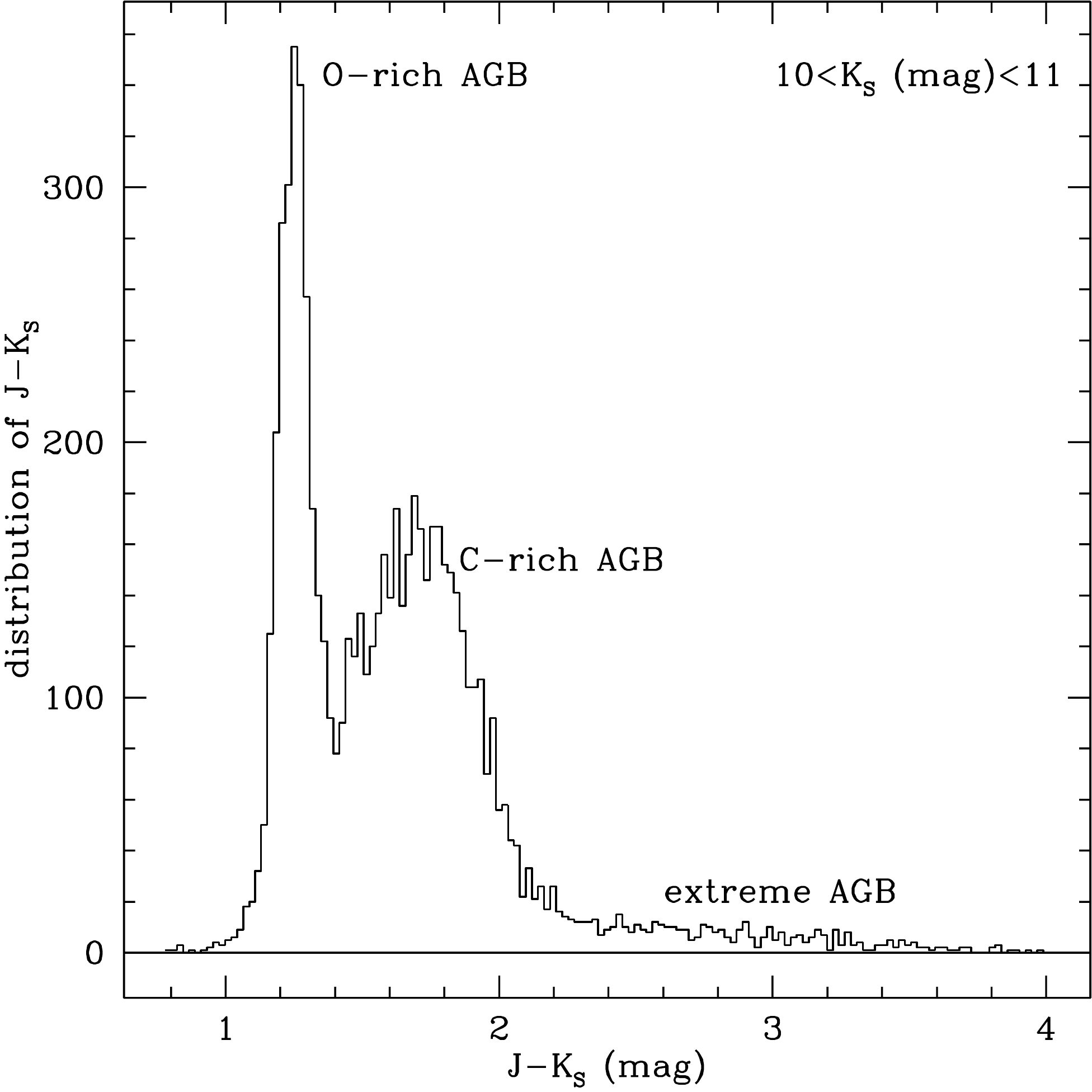}
   	\caption{Histogram of $J-K_{s}$ for variables with $10<K_{s}$\,(mag)$<11$ from Fig.~\ref{KJK_density}.}
   	\label{histo_JK_10_11}
\end{figure}

\begin{figure}
	\includegraphics[width=\columnwidth, height=7cm]{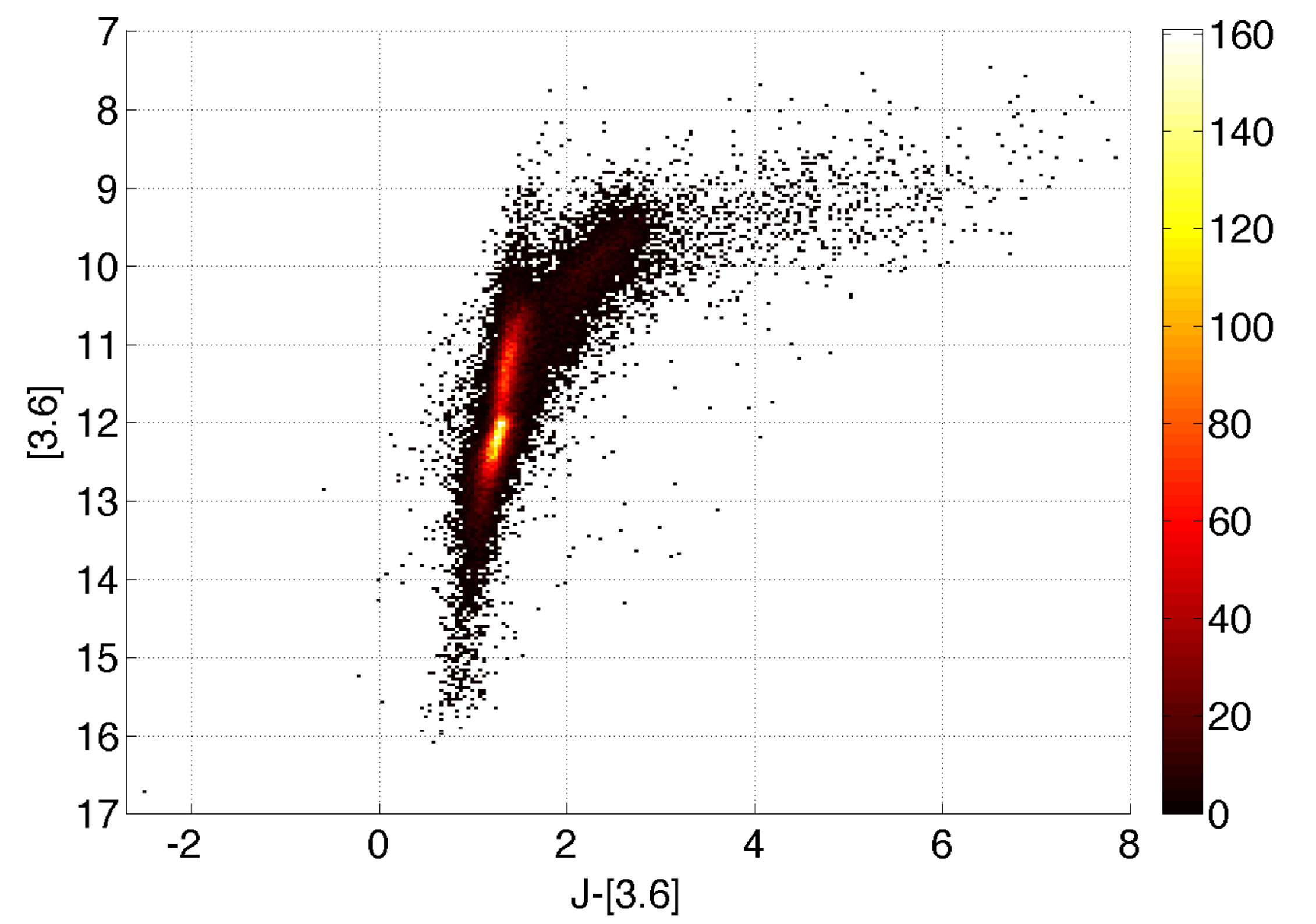}
   	\caption{Number density of the EROS-2 LPV candidates cross-matched with the IRAC catalog, in the $J-[3.6], [3.6]$ CMD. The CMD is divided in 250x250 bins with a bin size of 0.04x0.037 mag.}
   	\label{3.6J3.6_density}
\end{figure}

\begin{figure}
	\includegraphics[width=\columnwidth, height=6cm]{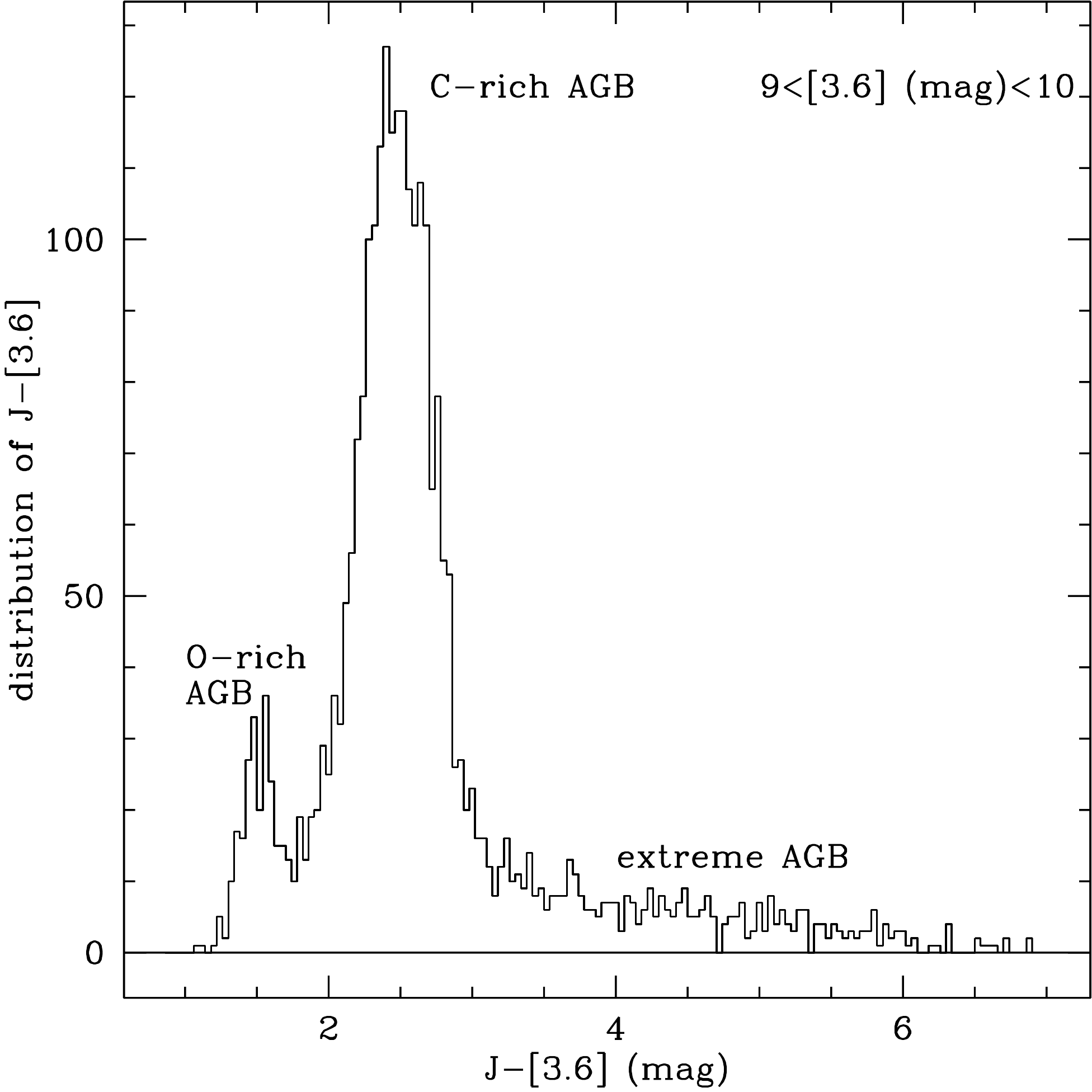}
   	\caption{Histogram of $J-[3.6]$ for variables with $9<[3.6]$\,(mag)$<10$ from Fig.~\ref{3.6J3.6_density}.}
   	\label{histo_J3_6_9_10}
\end{figure}

\begin{figure}
	\includegraphics[width=\columnwidth, height=4.5cm]{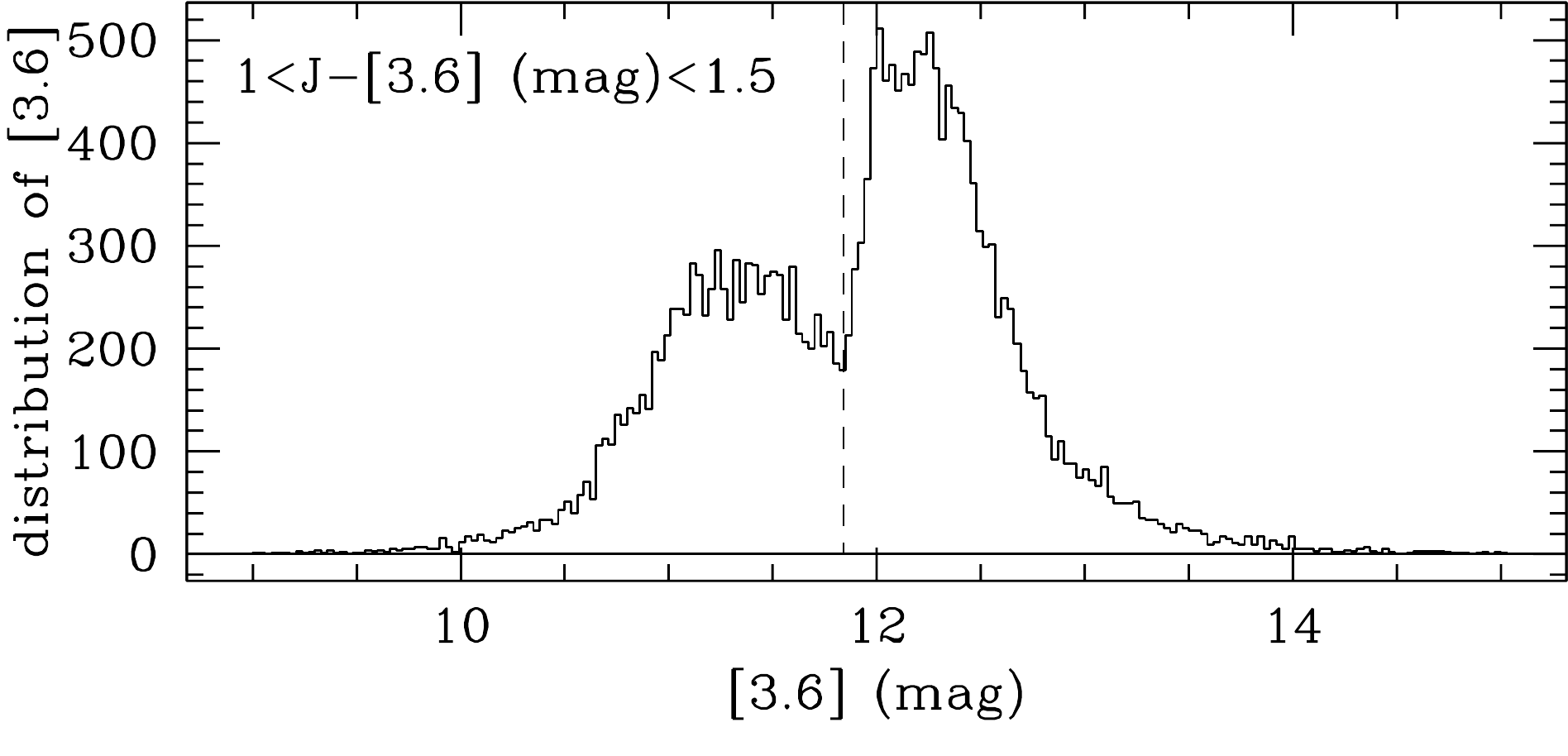}
   	\caption{Histogram of $[3.6]$ for variables with $1<J-[3.6]$\,(mag)$<1.5$ from Fig.~\ref{3.6J3.6_density}. Dashed line at $[3.6]=11.85$ indicates the magnitude of the TRGB.}
   	\label{histo_3_6_1_1.5}
\end{figure}

\begin{figure}
	\includegraphics[width=\columnwidth]{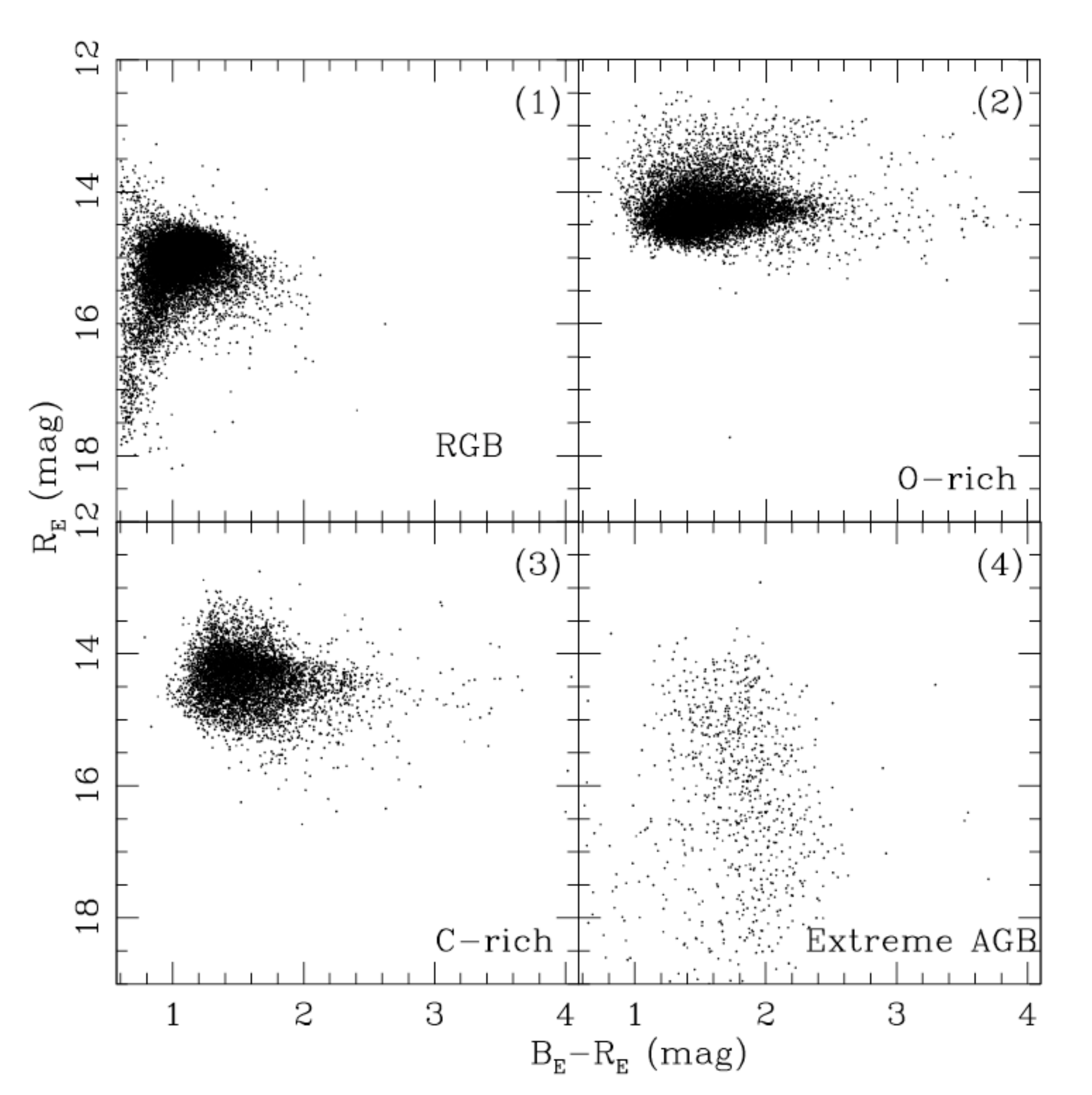}
   	\caption{($B_E-R_E, R_E$) CMD of the EROS-2 LPV candidates, according to the classification scheme into RGB, O-rich, C-rich, and extreme AGB stars defined in Sect.~\ref{Sect:Properties}.}
   	\label{CMD_EROS_SpType}
\end{figure}

We cross-matched the 43\,551 EROS-2 LPV candidates with the 2MASS All-Sky Catalog of Point Sources \citep{Cutrietal03} using the VizieR Services at "Centre de Donn\'ees astronomiques de Strasbourg". Using a search radius of 1 arcsec, we matched 41\,502 stars (95.3\% of the LPV candidates) in $J$ and $K_{s}$ bands, with a mean difference between EROS-2 and 2MASS coordinates of 0.31 arcsec, and a $\sigma_{d}$=0.16 arcsec. For three of these LPV candidates, the cross-matching ends with two possible 2MASS sources. In these cases, we kept only the source with the coordinates closest to those of our candidate and with the best 2MASS quality flag.\\
 We also cross-matched our candidates with the \textit{Spitzer} IRAC data at 3.6 microns, hereafter [3.6], from the SAGE database \citep{Meixneretal06}.  Within a search radius of 1 arcsec, 36\,847 stars (84.6\% of the LPV candidates) were cross-matched, with a mean distance between EROS-2 and IRAC coordinates of 0.33 arcsec, and a $\sigma_{d}$=0.18 arcsec. For three of these LPV candidates, we obtained two possible IRAC sources. In these cases, we kept only the source with the coordinates closest to those of our candidate.

Several methods exist to help us distinguish between O-rich and C-rich stars. Among these are the use of the reddening-free Wesenheit index $W_I$ combined with periods \citep{Soszynskietal05}, the use of either ($J-K, V-I$) color-color diagram, the ($W_{JK}, W_I$) diagram \citep{Soszynskietal09}, or the variability slope parameter $a_v$ versus color amplitude $amp_{VI}$ \citep{Wisniewskietal11}.
 Another way to distinguish between different spectral types in red giants, is to look at the $J-K$ color index, O-rich stars usually having a lower ($J-K$) value than C-rich \citep[cf.][]{NikolaevWeinberg00}.
 Fig.~\ref{KJK_density} shows the ($J-K_{s}$, $K_{s}$) Hess diagram for the LPV candidates. The two highest density zones are clearly separate at $K_{s}$=12 mag, which is the magnitude of the tip of the red giant branch (TRGB). We defined four zones in that diagram as follows:
\begin{itemize}
\item Zone 1 is mainly composed of RGB stars, although a contamination by faint AGB stars close to the TRGB is possible, as shown by \citet{Cionietal06}.
\item For Zone 2, we use \citet{Cionietal06} equations [6] and [7] to separate O-rich from C-rich AGB stars, respectively zones 2 and 3 in Fig.~\ref{KJK_density}, assuming a metallicity of Z=0.008 for the LMC.\\
 Zone 2, according to their models, contains O-rich thermally pulsating AGB (TP-AGB) stars, with presumably the brightest part ($K_{s}<10.6$ mag) populated by young M supergiants and luminous O-rich M stars as discussed in \citet{Blumetal06}.
\item Zone 3 contains C-rich TP-AGB stars, which evolved from the O-rich, though a mixture of these two populations is possible either by C-rich stars in zone 2 or by O-rich stars in zone 3. Fig.~\ref{histo_JK_10_11} shows the histogram of $J-K_{s}$ for variables from Fig.~\ref{KJK_density} with $10<K_{s}$\,(mag)$<11$. According to our  definitions of the zones, the highest peak is due to O-rich AGB stars from zone 2. For $J-K_{s}>1.4$ mag, C-rich AGB stars form the predominant population up to $J-K_{s}=2.1$ mag.
\item Zone 4 is composed of stars with $J-K_{s}>2.1$ mag, which tend to become fainter as they become redder, in constrast to O-rich and C-rich stars from zones 2 and 3. This population of stars called extreme AGB stars are known to be dusty evolved AGB stars \citep[see][]{Vijhetal09}.
\end{itemize}
 
 Owing to the presence of circumstellar dust from extreme AGB stars, we require mid-infrared data to allow the comparison of this population with less obscured O-rich and C-rich AGB stars. This is illustrated in \citet{Blumetal06} and \citet{ItaMatsunaga11}, where the use of \textit{Spitzer} SAGE data places extreme AGB stars on the brighter end of the C-rich branch, brighter stars being the redder ones. Combining 2MASS and SAGE data, \citet{Blumetal06} showed that the ($J-[3.6], [3.6]$) CMD is the most convenient to differentiate between O-rich, C-rich, and extreme AGB stars. In Fig.~\ref{3.6J3.6_density}, we plot the 36\,220 LPVs (83.1\% of the LPV candidates) for which we have both J and [3.6] data, as a ($J-[3.6], [3.6]$) Hess diagram. 
We find again the same features as in Fig.~\ref{KJK_density}. As expected, extreme AGB stars are generally brighter than C-rich AGB stars. Furthermore, C-rich stars are brighter than O-rich ones. This is clearly visible in the histogram in Fig.~\ref{histo_J3_6_9_10}, showing the distribution of $J-[3.6]$ for variables from Fig.~\ref{3.6J3.6_density} with $9<[3.6]$ (mag)$<10$, where C-rich stars are the main population, and the remaining O-rich stars are M supergiants. The signature of the TRGB is visible on the distribution of the $[3.6]$ magnitude for LPVs with $1<J-[3.6]$\,(mag)$<1.5$ in Fig.~\ref{histo_3_6_1_1.5} as a minimum of the distribution at $[3.6]=11.85$ mag.
As no models are available for ($J-[3.6], [3.6]$) diagram, we keep Cioni's models to select O-rich and C-rich TP-AGB stars from Fig.~\ref{KJK_density}, while extreme AGB stars are selected according to their $J-[3.6]$ color and RGB stars with respect to their [3.6] magnitude.
For the following study, LPV candidates are thus classified as follows: 
\begin{itemize}
\item RGB stars: LPV stars that are fainter than the TRGB, i.e $[3.6]>11.85$ mag. There are 17\,427 stars of this type in our sample.
\item Extreme AGB stars: those with $J-[3.6]>3.1$ mag, cf Fig.~\ref{histo_J3_6_9_10} and brighter than the TRGB in [3.6] band. These are the less numerous stars in our sample with 859 stars found.
\item O-rich stars belong to zone 2 from Fig.~\ref{KJK_density}, and not already identified as RGB or extreme AGB stars. These are the most numerous AGB stars in our sample with 10\,807 stars found.
\item C-rich stars belong to zone 3 from Fig.~\ref{KJK_density}, and not already identified as RGB or extreme AGB star. There are 6\,612 LPVs of this type in our sample.
\end{itemize}
Figure \ref{CMD_EROS_SpType} shows the distributions of these four groups of stars in the ($B_E-R_E, R_E$) CMDs. We can see that it would be much more difficult to separate the four groups with EROS-2 data alone. The TRGB for example is highly contaminated by O-rich and C-rich AGB stars in the ($B_E-R_E, R_E$) CMD. It is also impossible to disentangle O-rich from C-rich stars as they have the same $B_E-R_E$ color and $R_E$  magnitude range. Extreme AGB stars appear fainter than other stars, as expected for stars with thick circumstellar dust shells, but not redder. This diagram confirms the interest of using infrared data to distinguish LPVs with different abundances.

\subsection{Catalog}
\label{Sect:catalog}
\begin{table*}
\caption{Data sample of the LPV in the LMC from EROS-2. See Sect.~\ref{Sect:catalog} for column description.}  % title of Table
\label{table:2}      % is used to refer this table in the text
\centering                        % used for centering table
\resizebox{\textwidth }{!}{ %
\begin{tabular}{c c c c c r r r r r r r l l l l l}
\hline
ID & RA & DEC & $R_{E mean}$ & $(B_E-R_E)_{mean}$ & $Amp. R_E$ & $P1_{R_E}$ & $P2_{R_E}$ & $P3_{R_E}$ & $P4_{R_E}$ & $P5_{R_E}$ & NPer & $J$ & $K_{s}$ & $[3.6]$ & Spectral & Flag\\    % table heading 
~ EROS-2 & (J2000) & (J2000) & (mag) & (mag) & (mag) & (days)& (days)& (days) & (days)& (days) &  & (mag) & (mag) & (mag) & type & ~  \\
\hline
lm0020n24738 & 82.63144 & -69.35374 & 13.923 & 1.232 & 0.333 & 586.4 & 0 & 0 & 0 & 0 & 1 & 12.096 & 10.7 & 10.24 & C-rich & 0\\
lm0020n25074 & 82.53326 & -69.35603 & 16.832 & 0.647 & 0.258 & 0 & 0 & 0 & 0 & 0 & 0 & 999.9 & 999.9 & 15.213 & N/A& 0\\
lm0020n25567 & 82.59598 & -69.35854 & 14.826 & 1.082 & 0.124 & 30.2 & 0 & 0 & 0 & 0 & 1 & 13.33 & 12.241 & 12.14 & RGB& 0\\
lm0020n25716 & 82.26153 & -69.36008 & 14.762 & 1.303 & 0.218 & 46.8 & 404.6 & 2782.0 & 35.5 & 50.8 & 5 & 13.255 & 12.116 & 11.893 & RGB& 0\\
lm0020n26052 & 82.27126 & -69.36201 & 14.032 & 1.550 & 0.514 & 625.0 & 326.0 & 88.3 & 83.7 & 0 & 4 & 12.428 & 11.221 & 11.085 & O-rich& 1\\
lm0020n26446 & 82.65421 & -69.36325 & 15.084 & 0.924 & 0.090 & 31.1 & 202.2 & 33.5 & 0 & 0 & 3 & 13.82 & 12.719 & 12.604 & RGB& 0\\
lm0020n26983 & 82.36742 & -69.39571 & 17.100 & 0.694 & 0.511 & 0 & 0 & 0 & 0 & 0 & 0 & 999.9 & 999.9 & 999.9 & N/A& 0\\
lm0020n28519 & 82.36320 & -69.37551 & 15.062 & 1.428 & 1.836 & 0 & 0 & 0 & 0 & 0 & 0 & 13.232 & 11.527 & 11.11 & N/A& 92\\
\hline
\end{tabular}}
\end{table*}
In our catalog, we present the 43\,583 EROS-2 sources extracted as LPV candidates based on the Abbe value criterion and red clump color (cf. Sect.~\ref{Sect:AbbeLimitForLPVs}). Table~\ref{table:2} shows a sample of the catalog and contains the following information:\\
\textit{Column 1}: EROS-2 ID of the star according to the definitions by \citet{Derueetal02}.\\
\textit{Columns 2-3}: equatorial coordinates, RA and DEC, for the epoch J2000.\\
\textit{Column 4}: mean $R_E$ magnitude.\\
\textit{Column 5}: mean $B_E-R_E$ color.\\
\textit{Column 6}: $R_E$ band amplitude as a peak-to-peak value.\\
\textit{Columns 7-11}: up to five periods, in days, for the $R_E$ band, if found with all the three period search methods. In the following analysis we will refer to these periods.\\
\textit{Column 12}: the number of periods found and given in columns 7-11. The value is equal to 0 in case no common period was found by all the three period search methods.\\
\textit{Column 13}: 2MASS $J$ magnitude.\\
\textit{Column 14}: 2MASS $K_{s}$ magnitude.\\
\textit{Column 15}: \textit{Spitzer} $[3.6]$ magnitude.\\
\textit{Column 16}: the classification as defined in Sect~\ref{IRproperties}.\\
\textit{Column 17}: a flag value. Non-periodic RCB or DY Per type stars from \citet{Tisserandetal09} are flagged with value 91 or 92, respectively. Stars found to have harmonic periods among sequences D and E are flagged with the value 1. Flag is set to 0 in all the other cases.\\
The full catalog is available on-line at the CDS.

\subsection{Period-magnitude relations}
\subsubsection{Identification of sequences}
\label{SeqId}

\begin{figure}
	\includegraphics[width=\linewidth]{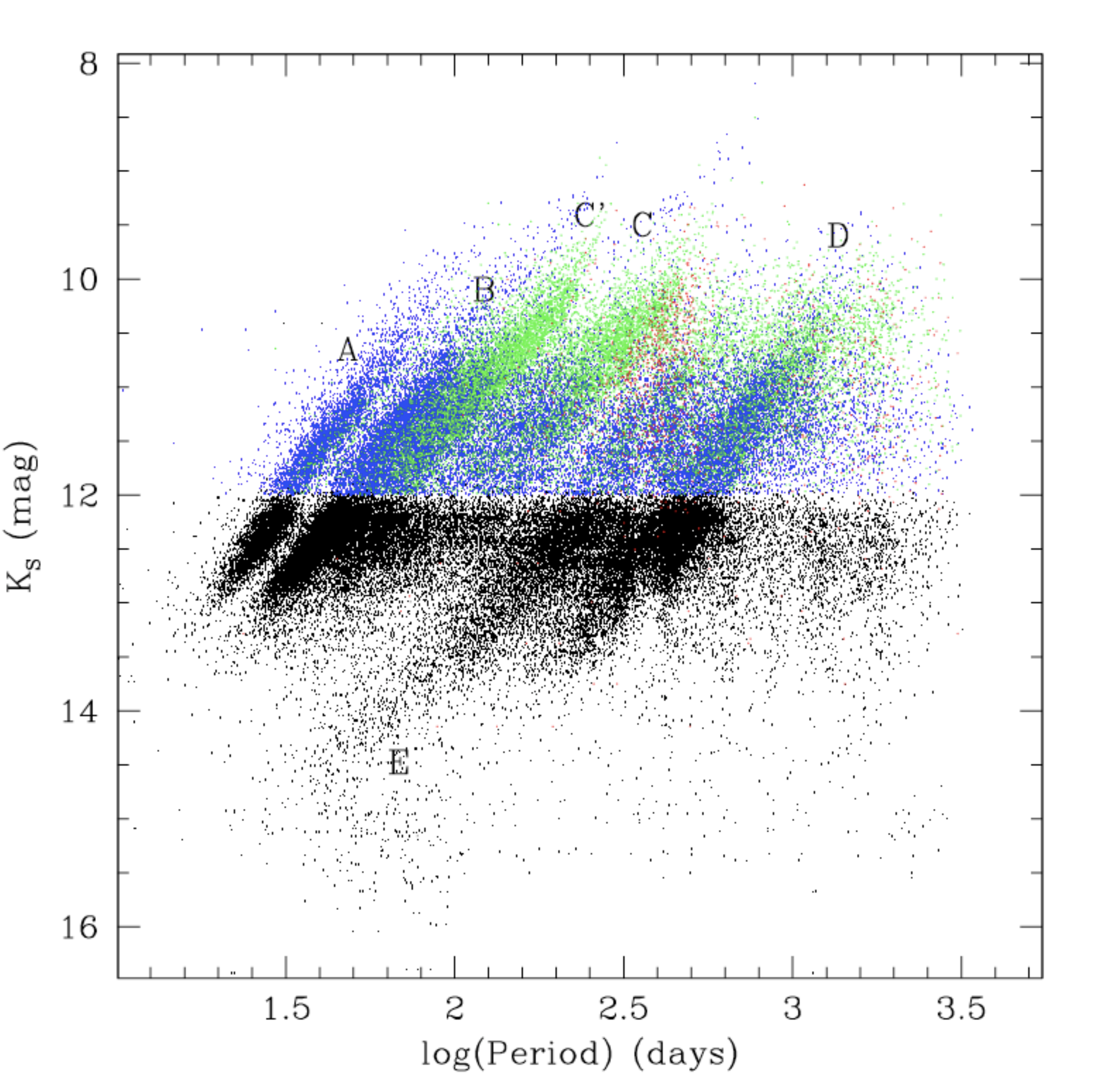}
   	\caption{(log $P, K_{s}$) diagram for all the periods of LPV candidates. The black points are RGB stars, blue points are O-rich AGB stars, green points are C-rich AGB stars, and red points, extreme AGB stars.}
   	\label{KlogPSpTyp}
\end{figure}

\begin{figure}
	\includegraphics[width=\linewidth]{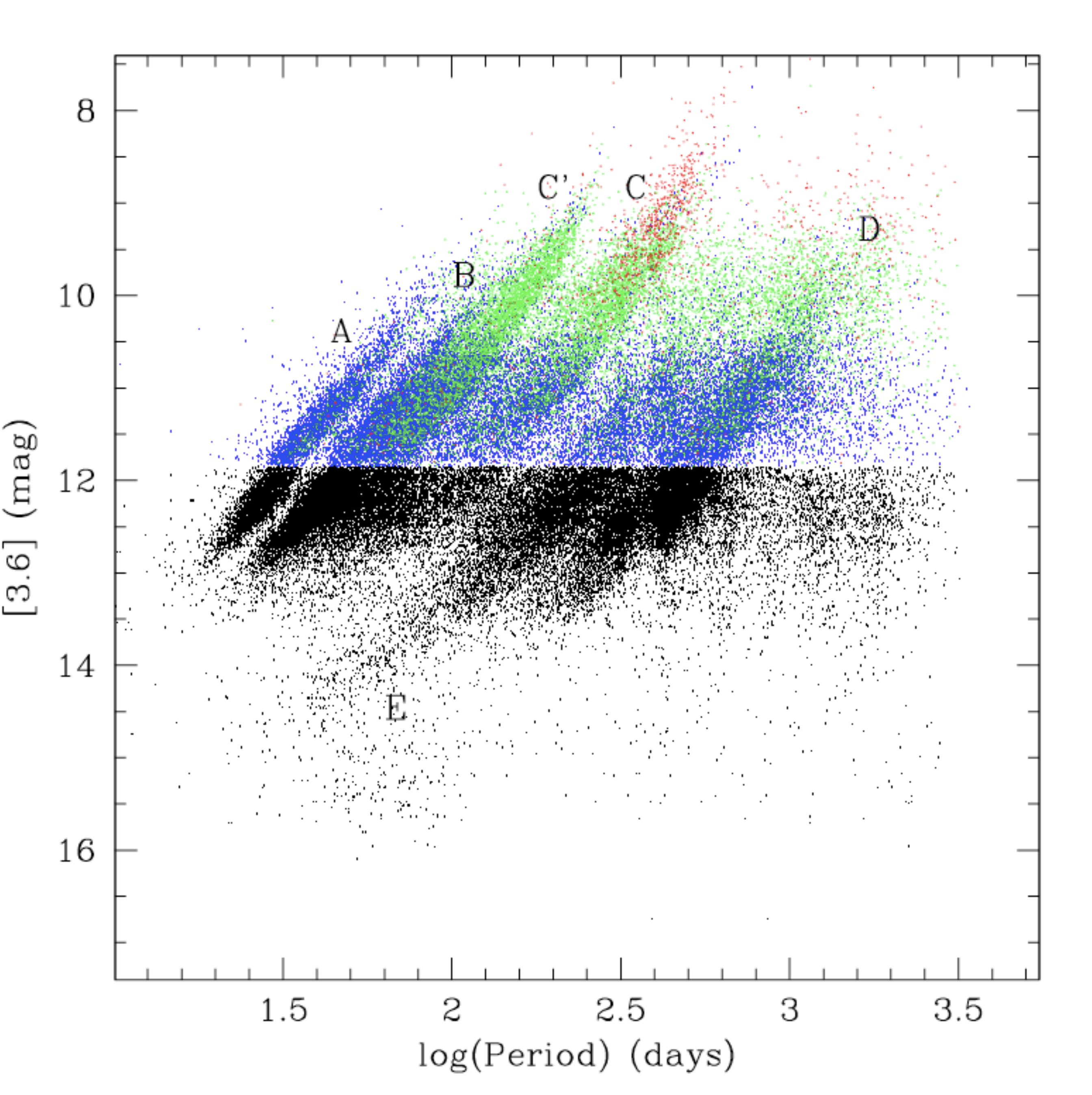}
   	\caption{Analogous to Fig.~\ref{KlogPSpTyp} for the $[3.6]$ magnitude versus log of the periods.}
   	\label{3_6logPSpTyp}
\end{figure}

\begin{figure}
	\includegraphics[width=\linewidth]{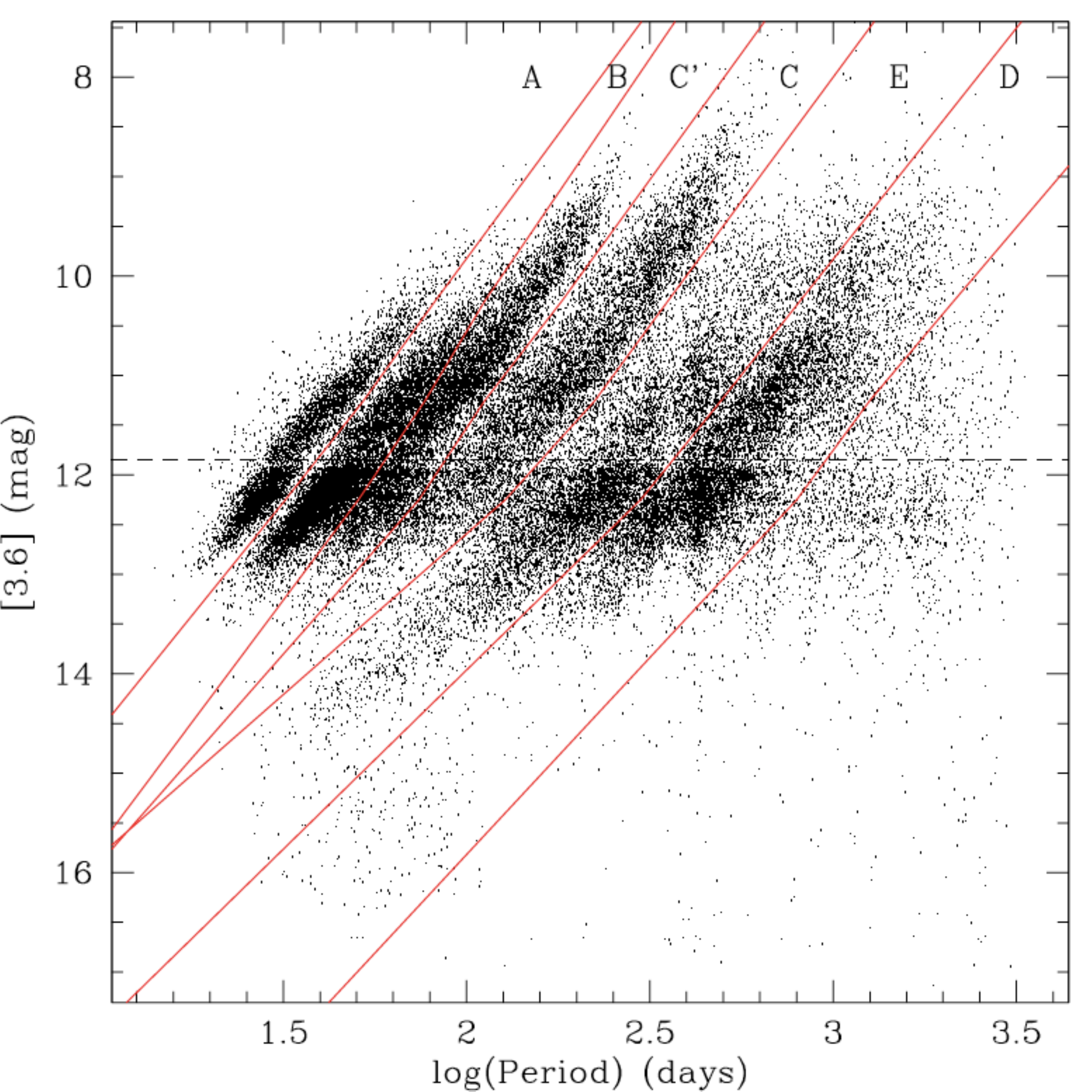}
   	\caption{(log $P, [3.6]$) diagram for all the periods selected for the LPV candidates. The dashed line indicates the $[3.6]$ magnitude of the TRGB. The red lines delimit the sequences as described in Sect~\ref{SeqId}.}
   	\label{PLSeq}
\end{figure}

\begin{figure}
	\includegraphics[width=\columnwidth]{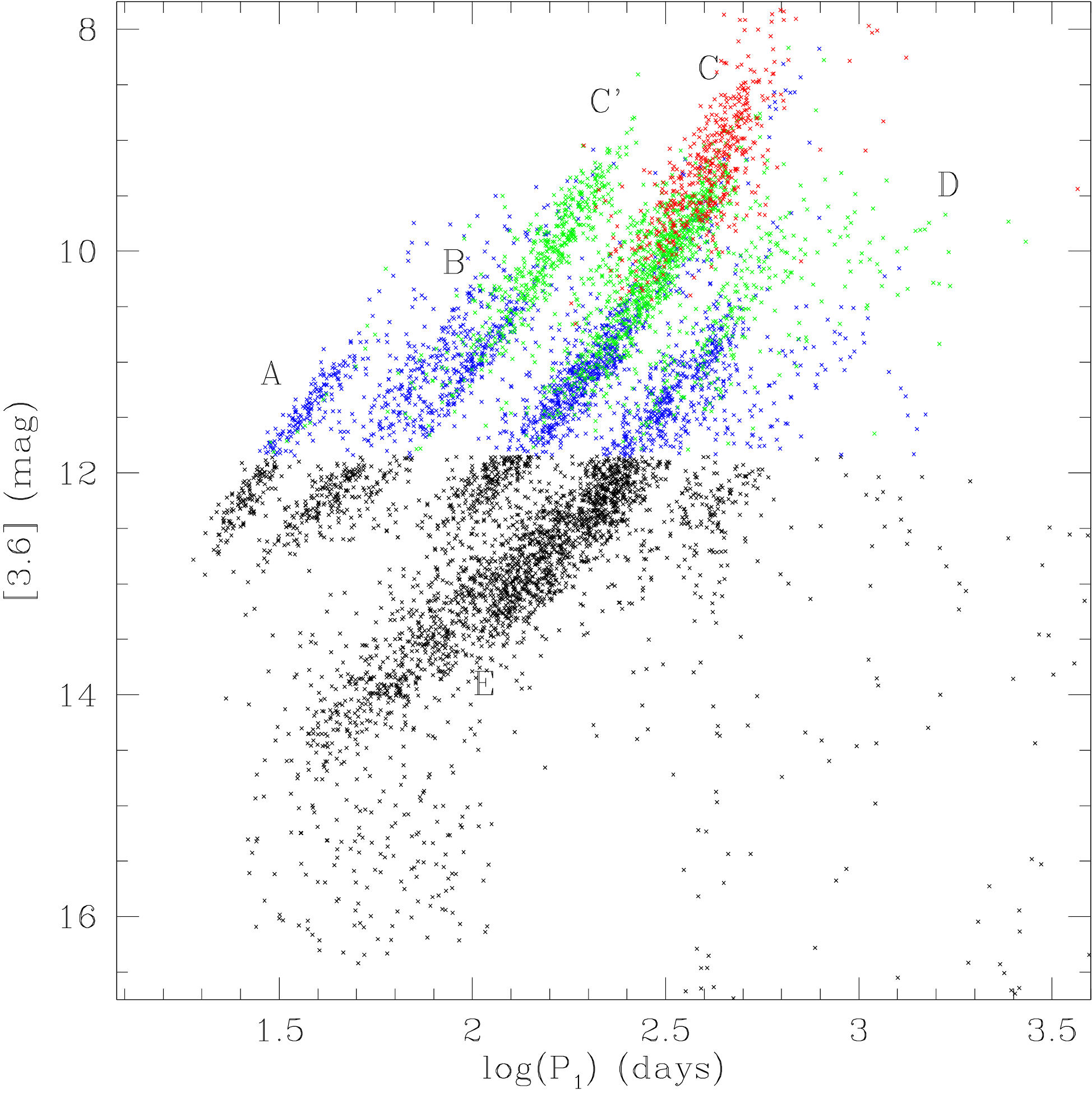}
   	\caption{(log $P, [3.6]$) diagram for LPVs with only one period identified. The black points are RGB stars, blue points are O-rich AGB stars, green points are C-rich AGB stars, and red points, extreme AGB stars.} 
   	\label{3.6_logP_1Per}
\end{figure}

\begin{figure}
	\includegraphics[width=\columnwidth]{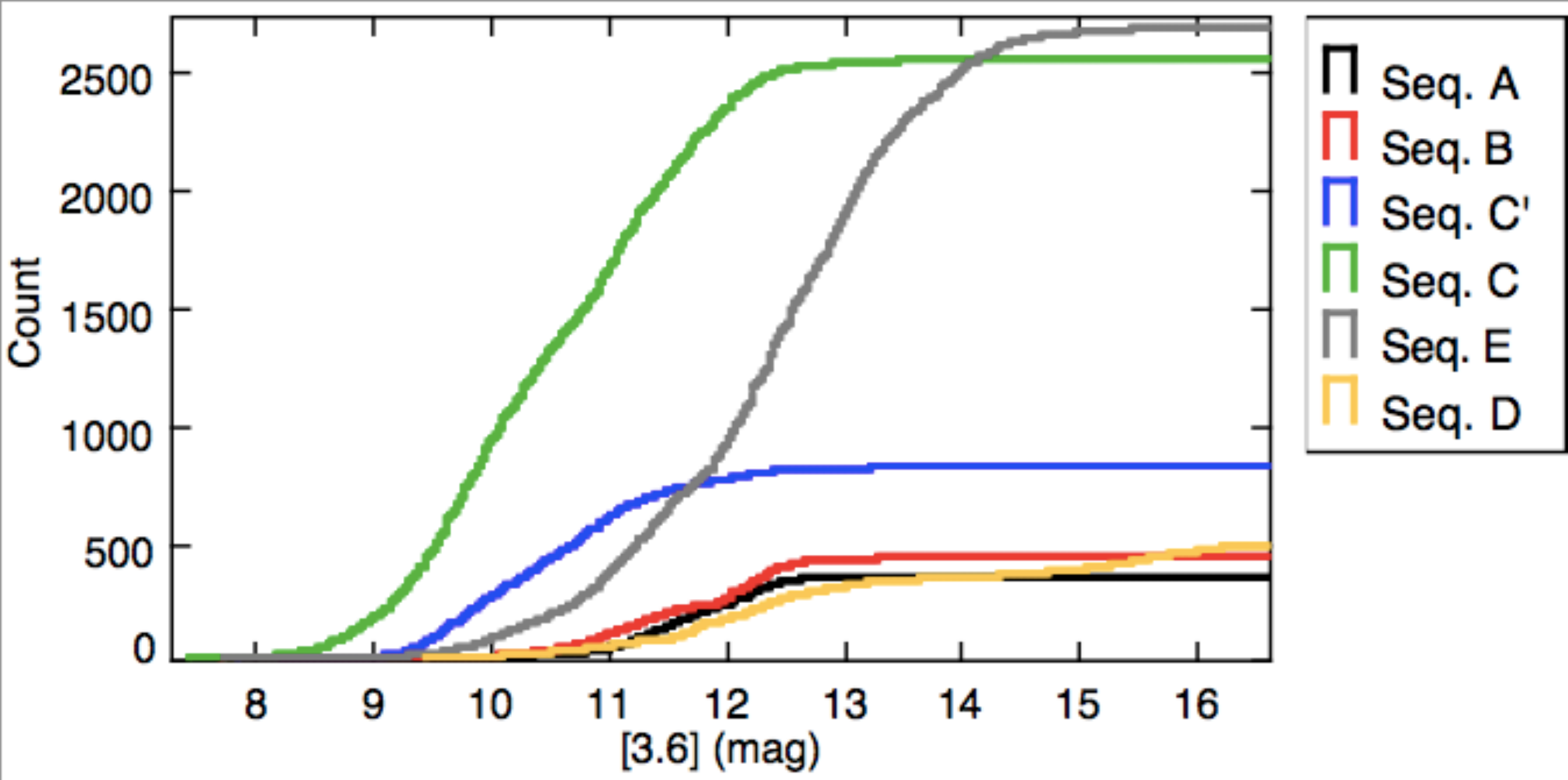}
   	\caption{Cumulative distribution for LPVs with only one period identified for the sequences in Fig.~\ref{3.6_logP_1Per}. Each line represents a sequence, which are, from the most populated (top line at $[3.6]=16.5$ mag) to the least populated (bottom line at $[3.6]=16.5$ mag): Seq. E, Seq. C, Seq. C', Seq. D, Seq. B, Seq. A.} 
   	\label{cumul_3.6_1Per}
\end{figure}

Figure~\ref{KlogPSpTyp} shows the ($log P, K_{s}$) diagram of our LPVs for all the periods, which means for example, that a variable with three periods will be represented by three points in the diagram. We color code the diagram according to the RGB, O-rich, C-rich, and extreme AGB star classification (see Sect.~\ref{IRproperties}). Six sequences are visible. Five of them, namely A, B, C, D, and E, have been studied in the seminal paper of \citet{Woodetal99}. \citet{Itaetal04} subdivide the sequence B into the B and C' sequences, using OGLE-II and SIRIUS data. We follow the same notation for our sequences as in these two works.

 Extreme AGB stars clearly produce a large scatter and a mix of population in the sequences. As shown in \citet{Riebeletal10} and \citet{ItaMatsunaga11}, the scatter in the period-magnitude relations is smaller when using mid-infrared data. Comparing 2MASS and SAGE data, \citet{Riebeletal10} showed in (log $P$, magnitude) diagrams that the sequences are more clearly defined when using the \textit{Spitzer} [3.6] band. The (log $P, [3.6]$) diagram in Fig.~\ref{3_6logPSpTyp} indeed, displays extreme AGB stars at the bright end, aligned with C-rich star sequences. Globally, the sequences are more easily differentiated using the $[3.6]$ band than the $K_{s}$ band. Hence, we use the (log $P, [3.6]$) diagram for further analysis.

In the (log $P, [3.6]$) diagram in Fig.~\ref{PLSeq}, we show all the periods found for our LPV candidates (including stars without defined spectral type). The signature of the TRGB is easily visible as a drop in star density at $[3.6]=11.85$ mag for the entire period range, as seen in other works \citep[cf.][]{KissBedding03}. 

We defined sequence boundaries by firstly connecting points in the valley of lowest density between the sequences. These points were identified as local minima of the distributions of log $P$ for three magnitude intervals: $12<[3.6]$(mag)$<12.5$, $11<[3.6]$(mag)$<11.5$, and $10<[3.6]$(mag)$<10.5$. Secondly, as  the density of points is too low at the bright and faint magnitudes we visually add points at $[3.6]=8$ mag and $[3.6]=14$ mag to adjust the shape of the boundaries to the sequences. We performed two least squares fits for each of these six sequences, one for the region of the diagram brighter and one for the region fainter than the TRGB. The parameters of the equations we obtained are summarized in Table~\ref{table:2bis}. Except for sequence A, all the sequences show a steeper slope for AGB stars than for RGB ones. 

\begin{table}
\caption{Linear fit parameters for the sequences from Fig.~\ref{PLSeq} with respect to the TRGB.}             % title of Table
\label{table:2bis}      % is used to refer this table in the text
\centering                        % used for centering table
\begin{tabular}{|c|c|c|c|c|} 
\hline
Sequence & \multicolumn{2}{c|}{RGB} & \multicolumn{2}{c|}{AGB} \\
~  & Slope & Intercept & Slope & Intercept ~  \\\hline
A & -5.64 & 20.31 & -4.00 & 17.77\\ 
B & -3.52 & 17.93 & -4.70 & 19.76\\ 
C' & -3.77 & 18.95 & -4.82 & 20.64\\ 
C & -2.41 & 17.02 & -4.33 & 20.72\\ 
E & -2.77 & 18.72 & -3.61 & 20.31\\
D & -3.05 & 20.36 & -3.35 & 20.71\\
\hline
\end{tabular}
\end{table}

\begin{table*}
\caption{LPV period distribution among the sequences. In the cases of multi-periodic LPVs, a sequence is attributed to each period found.}
\label{table:3}      % is used to refer this table in the text
\centering                        % used for centering table
\begin{tabular}{|c|r|c|r|r|r|r|r|r|} 
\hline
Number & \multicolumn{2}{c|}{Number of LPVs} & \multicolumn{6}{c|}{Distribution of periods in} \\
~  of periods & in total & in $[3.6]-\log P$ diag. & Seq. A & Seq. B & Seq. C' & Seq. C & Seq. E & Seq. D~  \\
\hline
0 & 6\,040 (13.86\%) & - & - & - & - & - & - & -\\ 
1 & 9\,666 (22.18\%) & 7\,532 & 347 & 428 & 818 & 2\,549 & 2\,675 & 478\\ 
2 & 9\,904 (22.72\%) & 8\,496 & 1\,206 & 2\,255 & 3\,054 & 2\,790 & 2\,024 & 4\,845\\ 
3 & 8\,595 (19.72\%) & 7\,844 & 2\,173 & 4\,451 & 3\,645 & 2\,805 & 3\,430 & 6\,140\\ 
4 & 5\,845 (13.41\%) & 5\,513 & 2\,220 & 4\,930 & 3\,253 & 2\,217 & 3\,637 & 5\,038\\
5 & 3\,533 (08.11\%) & 3\,403 & 1\,795 & 3\,921 & 2\,375 & 1\,578 & 3\,071 & 3\,668\\ 
\hline
\end{tabular}
\end{table*}

\subsubsection{Cleaning of harmonics}
\label{cleaning}
In the following sections, we analyze mono and double periodic LPVs. Since light curves of ellipsoidal variables have minima of alternating depth, our period search could lead to the selection of harmonic periods. We handle these harmonics for double-periodic LPVs that are potential binaries, as follows: if a LPV has one of its periods in sequence D and the other in sequence E, with $P_{D}=2P_{E}$$\pm$10\% we remove the sequence D period and attribute only the sequence E period to the LPV. This means that we have binaries with minima of alternating depth and binaries with minima of similar depth in sequence E. A second step was to study all the LPVs that remain in the sequence D. We looked by eye at the raw and phase-folded light curves. If it appears to be typical of an ellipsoidal variable, we attribute a period $P=P_{D}/2$ as their only period. For LPVs with more than two periods found, we attribute a flag in the catalog if two of these periods are multiple by a factor of two, with one period in sequence D and any other one in sequence E.

\subsubsection{Mono-periodic LPVs}
\label{MonoPer}
Data in the $[3.6]$ band were available for 7\,532 out of 9\,666 LPVs classified as mono-periodic. Fig.~\ref{3.6_logP_1Per} displays the (log $P, [3.6]$) diagram for these stars.
The most populated sequence for the mono-periodic LPVs appears to be sequence E, as shown in the cumulative distribution of Fig.~\ref{cumul_3.6_1Per}. Sequence E is known to contain ellipsoidal binaries, effectively found with only one period, but with half the value of the orbital period because of the methods used for the period search, as shown in studies of binary systems \citep[see][]{Soszynskietal04, Derekasetal06}. As visible in Fig.~\ref{3.6_logP_1Per}, variables from this sequence are present at all magnitudes. Nevertheless, Fig.~\ref{cumul_3.6_1Per} shows that most of the sequence E variables are fainter than the TRGB ($[3.6]=11.85$ mag) in contrast to all the other sequences.
The second most populated sequence, with more than 25\% of the mono-periodic LPV sample, is the sequence C. The stars on this sequence are radially pulsating in the fundamental mode, the so-called Miras. Except for a very few cases in sequences C' and D, sequence C is the only one containing extreme AGB stars. 

Some variables with only one period are distributed over sequences C', B, and A, populated by overtone pulsators. Sequence C' is populated by first overtone pulsators (cf. \citet{WoodSebo96}, \citet{Itaetal04}). Sequences A and B are expected to be populated with semi-regular small amplitude red-giant variables pulsating in their second or third overtones. The nature of these variables is not easy to determine, because available models of low-mass (in the range 0.8 to 1.5 $M_{\sun}$) pulsating stars in overtones modes are unable to reproduce the sequences observed \citep[cf.][]{Woodetal99}.
 These sequences are probably composed of a mix of classical semi-regular variables and OSARGs that may require different physical processes to explain their variability \citep{Soszynskietal07}. 

Only a few stars are visible in sequence D, which usually contains LSP variables, that have more than one period by definition. Visual inspection of our LPV candidate light curves in this sequence reveals that it contains mostly stars that exhibit a LSP feature with a second period that is too short (i.e. shorter than 10 days) to be detected by our period search. Some other types such as irregular variables with a long-term trend, and peculiar stars exhibiting long period variability with an amplitude variation, probably due to spotted stars (cf. \citet{Marquetteetal08}), are also found in this sequence.

\subsubsection{Double-periodic LPVs}
\label{DoublePEr}
\begin{figure}
	\includegraphics[width=\columnwidth]{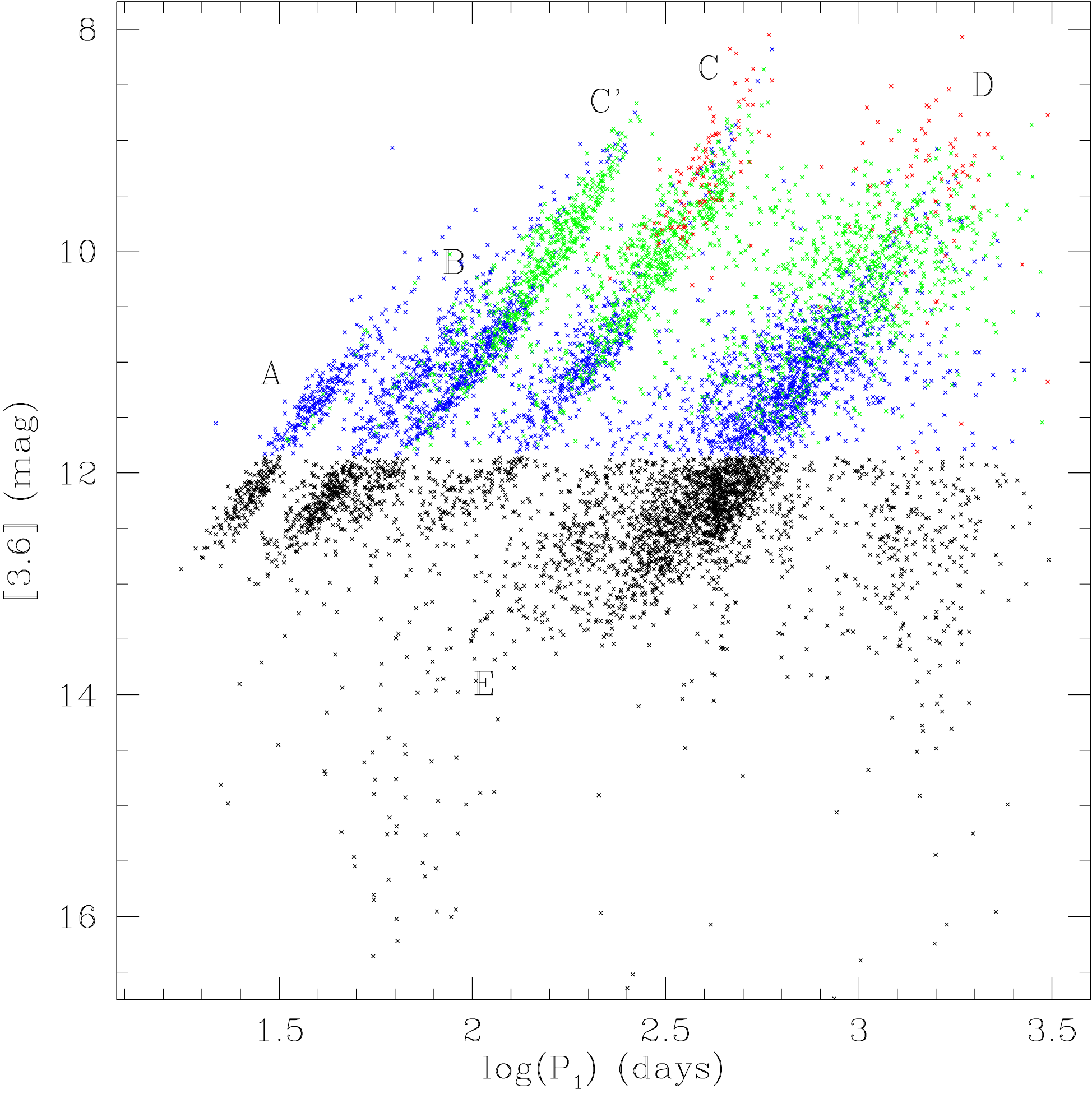}
   	\caption{Log of the first period versus $[3.6]$ diagram for LPVs with two identified periods. Colors are as in Fig.~\ref{KlogPSpTyp}.}
   	\label{3.6_logP_2PerP1}
\end{figure}

\begin{figure}
	\includegraphics[width=\columnwidth]{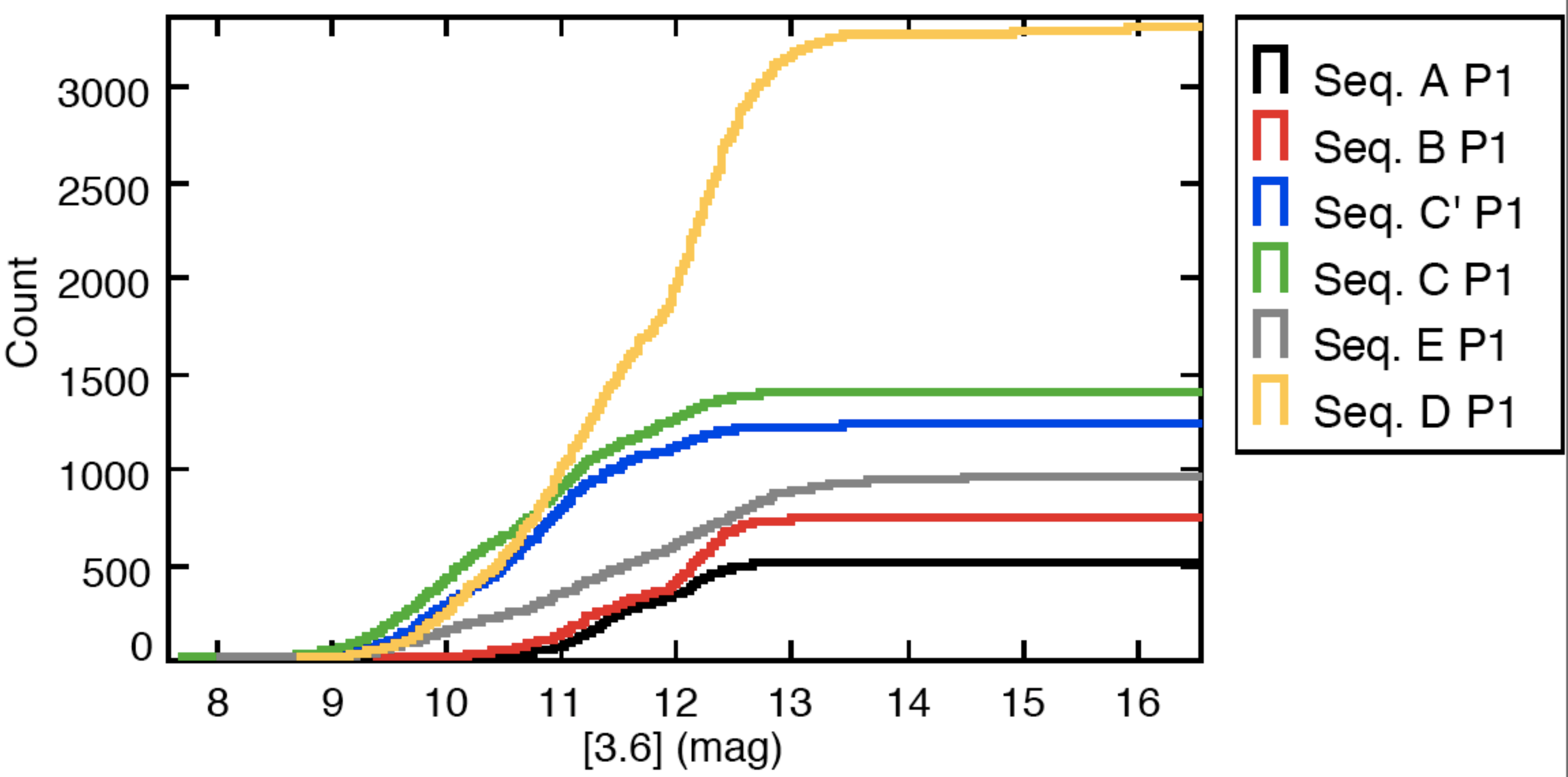}
	\includegraphics[width=\columnwidth]{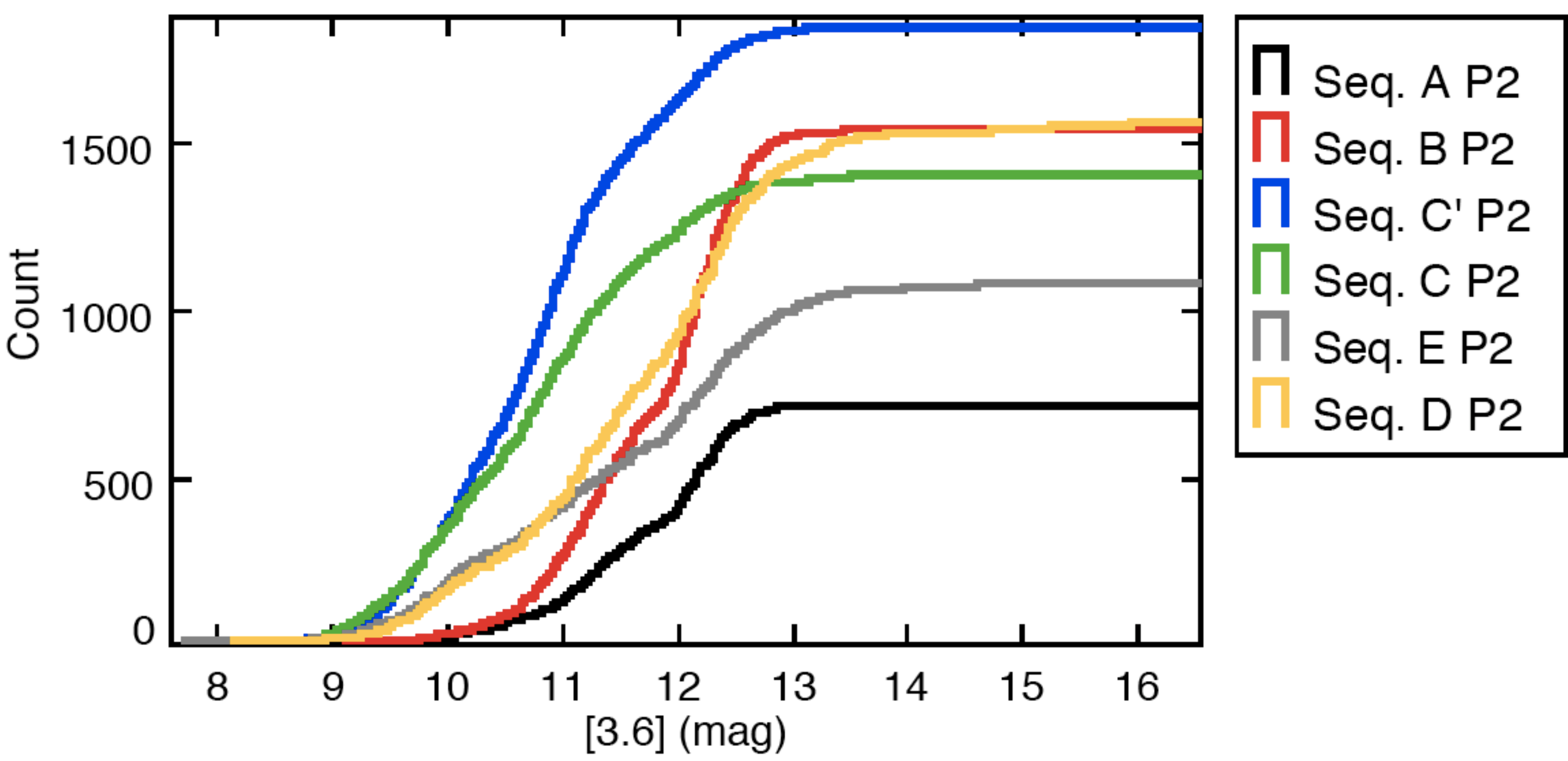}
   	\caption{Cumulative distribution for double-periodic LPVs for the sequences in Fig.~\ref{3.6_logP_1Per}. Each line represents a sequence, which are, from the most populated (top line at [3.6]=16.5mag) to the less populated (bottom line at [3.6]=16.5mag): 
	For the first period (top panel): Seq. D, Seq. C, Seq. C', Seq. E, Seq. B, Seq. A. 
	For the second period (bottom panel): Seq. C', Seq. D, Seq. B, Seq. C, Seq. E, Seq. A.} 
   	\label{cumul_3.6_2Per}
\end{figure}

\begin{figure}
	\includegraphics[width=\columnwidth]{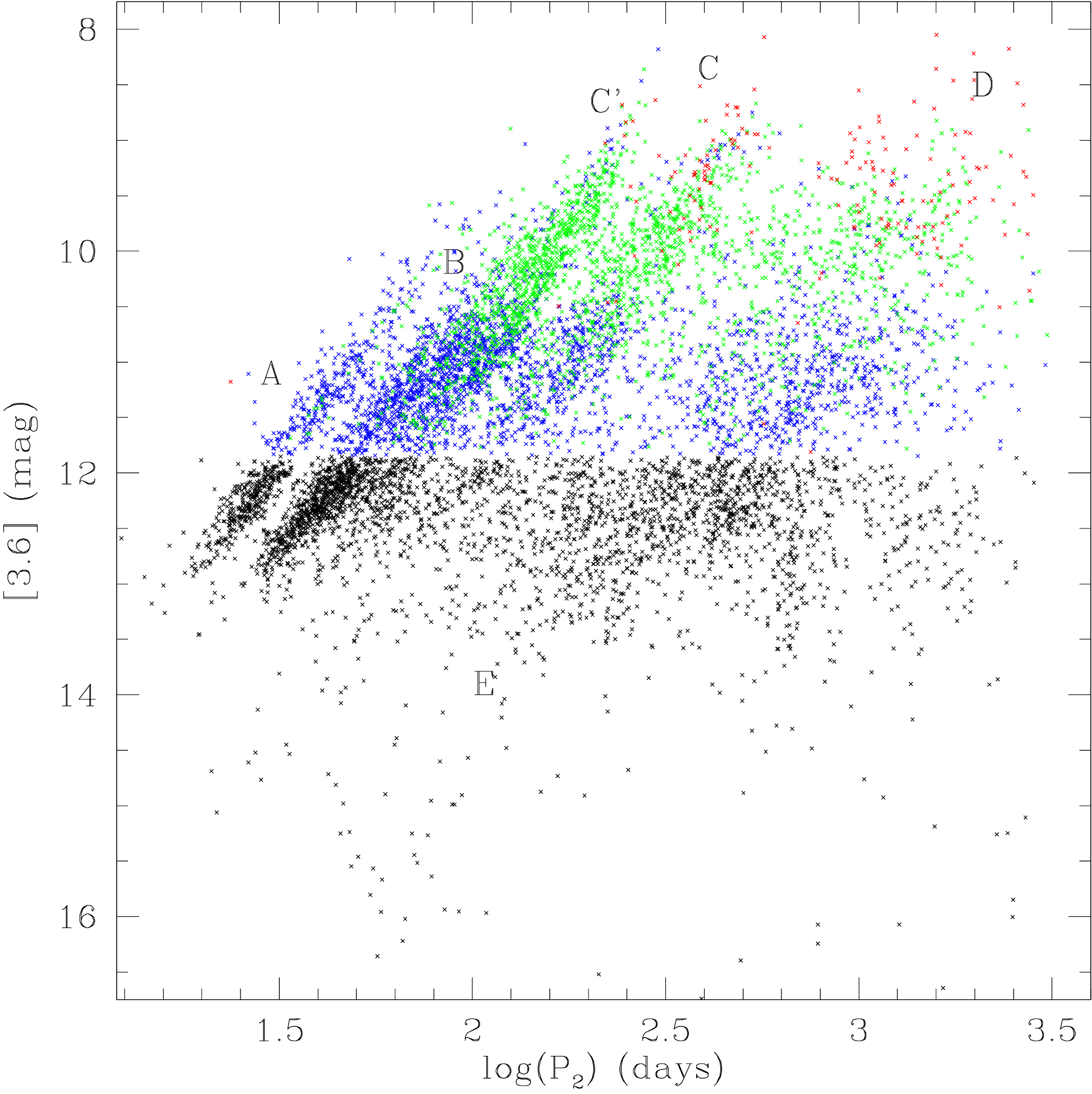}
   	\caption{Log of the second period versus $[3.6]$ diagram for double-periodic LPVs. Colors are as in Fig.~\ref{KlogPSpTyp}.}
   	\label{3.6_logP_2PerP2}
\end{figure}

\begin{figure}
	\includegraphics[width=\columnwidth]{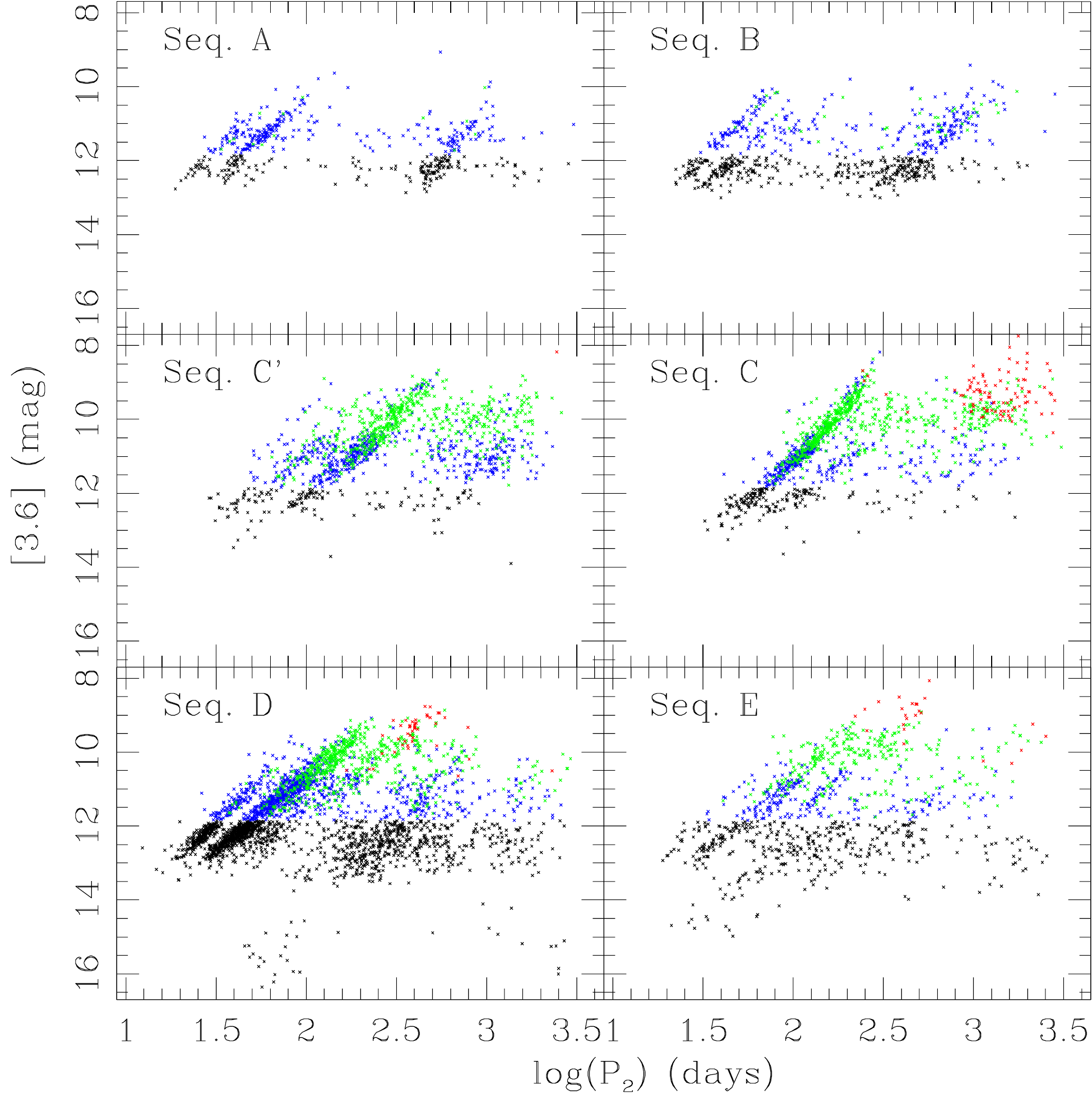}
   	\caption{Log of the second period versus $[3.6]$ diagram for double-periodic LPVs, according to the sequence in which the first period was found.  Colors are as in Fig.~\ref{KlogPSpTyp}.}
   	\label{3.6_logP_2PerP2_P1seq}
\end{figure}

The most numerous subtype in our sample are double-periodic LPVs, making up one quarter of LPV candidates (cf. Table~\ref{table:3}). For each of these stars, we assign two sequences with respect to the position in the (log $P, [3.6]$) diagram of the periods found.
 Figure ~\ref{3.6_logP_2PerP1} shows the (log $P, [3.6]$) diagram of the first period found, for the double-periodic LPVs. One can see the few extreme AGB stars in sequence C relative to the mono-periodic case (cf. Fig.~\ref{3.6_logP_1Per}). Figure \ref{cumul_3.6_2Per} (top panel) shows the cumulative distribution for each sequence in the [3.6] band. It shows that  the majority of the population consists of LPVs in sequence D, supposedly LSPs, for which almost half of these stars are fainter than the TRGB. Double-periodic stars with their first period in fundamental (sequence C) or low-order overtone mode (sequence C') are the next most numerous variables. There is still a noticeable population of stars in sequence E. Semi-regular variables from sequences A and B are more numerous than for the mono-periodic case ($+44\%$ of LPVs for sequence A and $+69\%$ for sequence B), but remain the least populated sequences.

Figure~\ref{3.6_logP_2PerP2} shows the (log $P, [3.6]$) diagram of all the second periods found. In Fig.~\ref{3.6_logP_2PerP2_P1seq}, we show six (log $P, [3.6]$) panels for the second period according to where the first period was found. This allows us to see the origins of the sequences of the (log $P, [3.6]$) diagram in Fig.~\ref{3.6_logP_2PerP2}. Sequence C' contains the largest number of LPVs (cf. bottom panel of Fig.~\ref{cumul_3.6_2Per}), most of them having their first period in sequence D, as noted first by \citet{Woodetal99}. There is also a significant number of stars in sequence C' coming from sequences C and E. One can see that most LPVs with their first period in sequence C' (resp. C) have their second period in sequence C (resp. C') (cf. middle panel in Fig.~\ref{3.6_logP_2PerP2_P1seq}).
 This exchange between these two sequences from the first to the second period, suggests that they are semi-regular variables.
Sequence D is one of the most populated, mainly because of LPVs with their first period in either sequence A or B. There are also stars coming from sequences C' and C, in particular C-rich and extreme AGB ones.
 Sequence B, and to a lesser extend sequence A, in Fig.~\ref{3.6_logP_2PerP2_P1seq} are dominated by stars whose first period places them in sequence D. There is also a swap between sequences A and B from the first to the second period. It is more noticeable for semi-regulars with their first period in sequence A for which half have their second period in sequence B, than for first-period sequence B semi-regulars for which only 18\% have their second period in sequence A.
Sequence E variables represent 10\% of the double-periodic LPVs. Half of them have their first period in sequence D, the other half being distributed among the other sequences.

\begin{table*}
\caption{Basic parameters of the sequences of Fig.~\ref{3_6logPSpTyp}. We assigned one sequence per star according to its first period found.}  % title of Table
\label{table4}      % is used to refer this table in the text
\centering                        % used for centering table
\begin{tabular}{l r l l l l l l}
\hline
Seq. & Number & log($Amp_{mean}$) & log($Amp_{mean}$) & log($Amp_{mean}$) & log($Amp_{mean}$) & log($Amp_{mean}$) & log ($P_1$)\\    % table heading 
~ & of stars & All & RGB & O-rich & C-rich & Extr. AGB & (days)-All ~ \\
\hline
A & 1\,982 & -0.939 & -1.022 & -0.888 & -0.303 & - & 1.565\\
B & 3\,438 & -0.800 & -0.865 & -0.727 & -0.679 & 0.071 & 1.730\\
C' & 3\,574 & -0.412 & -0.630 & -0.403 & -0.356 & 0.226 & 2.059\\
C & 5\,186 & -0.072 & -0.337 & -0.109 & -0.176 & 0.331 & 2.383\\
E & 5\,271 & -0.546 & -0.768 & -0.475 & -0.342 & 0.387 & 2.475\\
D & 11\,933 & -0.589 & -0.795 & -0.550 & -0.378 & 0.407 & 2.792\\
\hline
\end{tabular}
\end{table*}

\subsection{Period-amplitude relations}
\begin{figure}
	\includegraphics[width=\columnwidth]{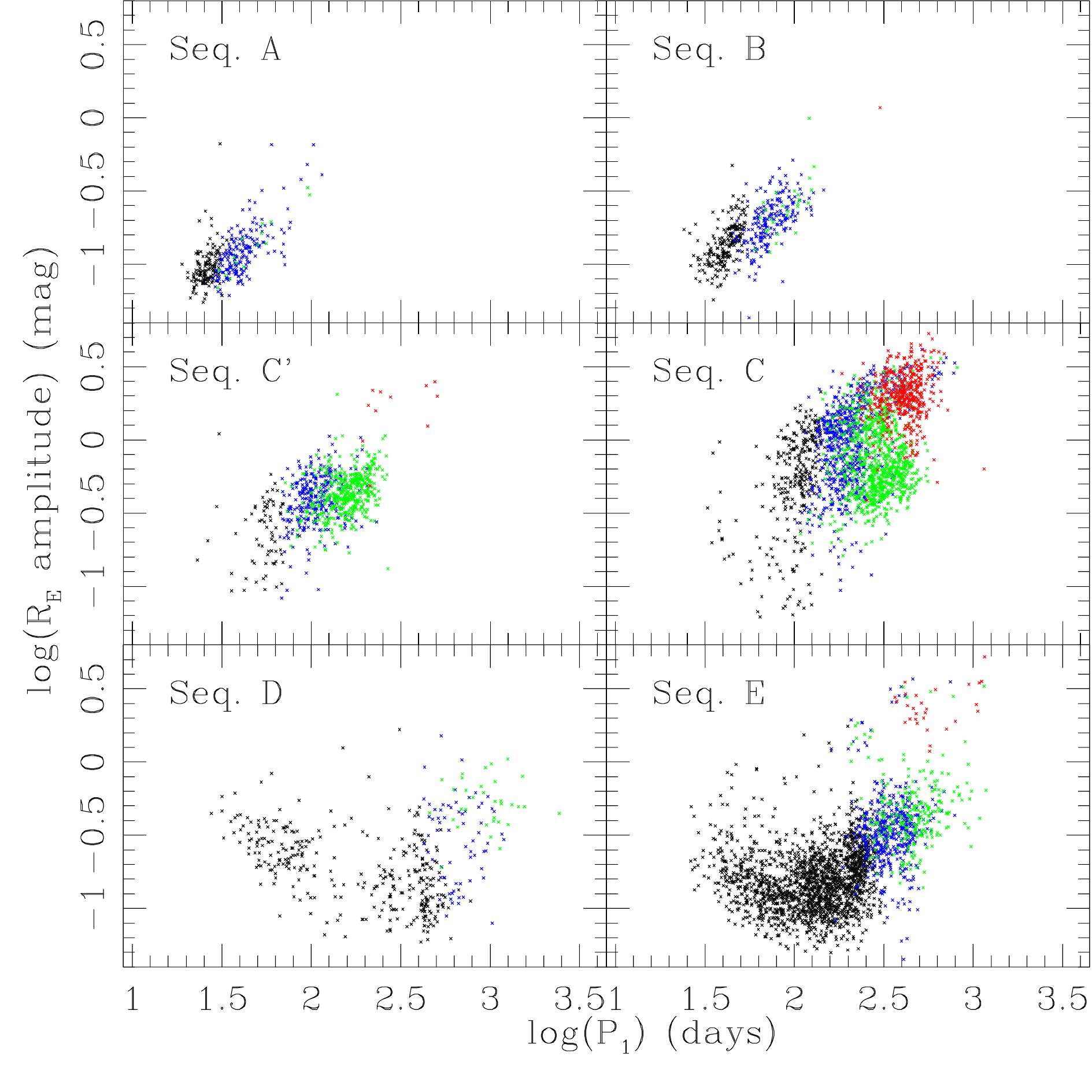}
   	\caption{Period-amplitude diagram, on a log scale, for mono-periodic LPVs according to the sequences of Fig.~\ref{3.6_logP_1Per}. Colors are as in Fig.~\ref{KlogPSpTyp}.}
   	\label{PerAmp1Per}
\end{figure}

Figure ~\ref{PerAmp1Per} shows the period-amplitude diagram for mono-periodic LPVs for each of the sequences defined in Sect.~\ref{SeqId}. Amplitudes were computed as a peak-to-peak difference in the light curves from the R band. From fundamental to high overtone mode pulsators, one can see that the mean amplitude decreases (cf. Table~\ref{table4}), in accordance with theory, which predicts that fundamental mode pulsators have larger amplitudes than overtones ones. Within each sequence, we also observe an increase in amplitude with period depending on the type of LPV as defined at the end of Sect.~\ref{IRproperties}. The RGB stars have the shortest periods and on average the smallest amplitudes, whereas extreme AGB stars, when present, have the longest periods and largest amplitudes (cf. Table~\ref{table4}). The O-rich and C-rich LPVs have intermediate periods and amplitudes, O-rich LPVs having shorter periods than C-rich ones. However, it is impossible to visually distinguish these two types in terms of their amplitudes.

 Apart from a few cases, the amplitude for stars from sequences A and B remains below 0.3 mag ($\log Amp=-0.5$). As previously mentioned in Sect.~\ref{MonoPer}, a mix of SRVs and OSARGs is possible in these sequences. Sequence C' contains mainly O-rich and C-rich LPVs with amplitudes between 0.2 ($\log Amp=-0.7$) and 0.8 mag ($\log Amp=-0.1$), with a few extreme AGB stars at amplitudes larger than 1 mag.

Sequence C contains RGB to extreme AGB stars with most of the amplitudes between 0.3 ($\log Amp=-0.5$) and 5 magnitudes ($\log Amp=0.7$).

The RGB stars form the main population of sequence E and have a low mean amplitude. No clear linear correlation between period and amplitude is visible. However, the lower envelope formed by the distribution of RGBs seems to show a decrease in amplitude with period up to $\log P=2$ followed by an increase in amplitude toward higher periods. At longer periods, $\log P>2.5$, an increase in amplitude is visible for O-rich and C-rich LPVs, extreme AGB stars being again among the largest amplitude LPVs.

The shape of sequence D appears very similar to that of sequence E. It could contain two populations: LSPs made of O-rich, C-rich, and longer period RGB stars, and stars that indeed could be a part of the faint end of sequence E, visible as a group of stars with $[3.6]>15$ mag and $1.5<\log P<2$ in Fig.~\ref{3.6_logP_1Per}.
\begin{figure}
	\includegraphics[width=\columnwidth]{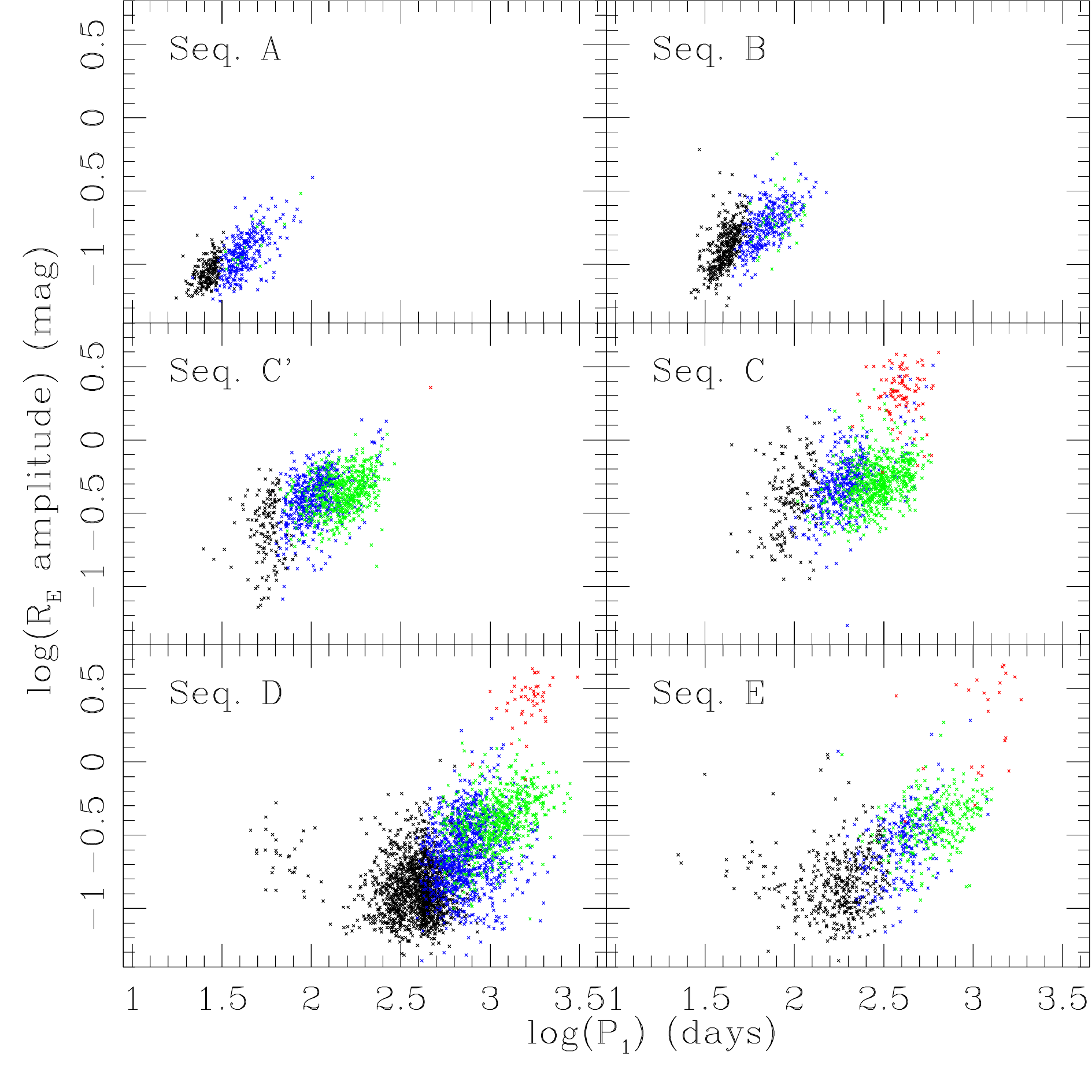}
   	\caption{Period-amplitude diagram, in log scale, for the first period of double-periodic LPVs according to the sequences of Fig.~\ref{3.6_logP_1Per}. }
   	\label{PerAmp2PerP1}
\end{figure}

\begin{figure}
	\includegraphics[width=\columnwidth]{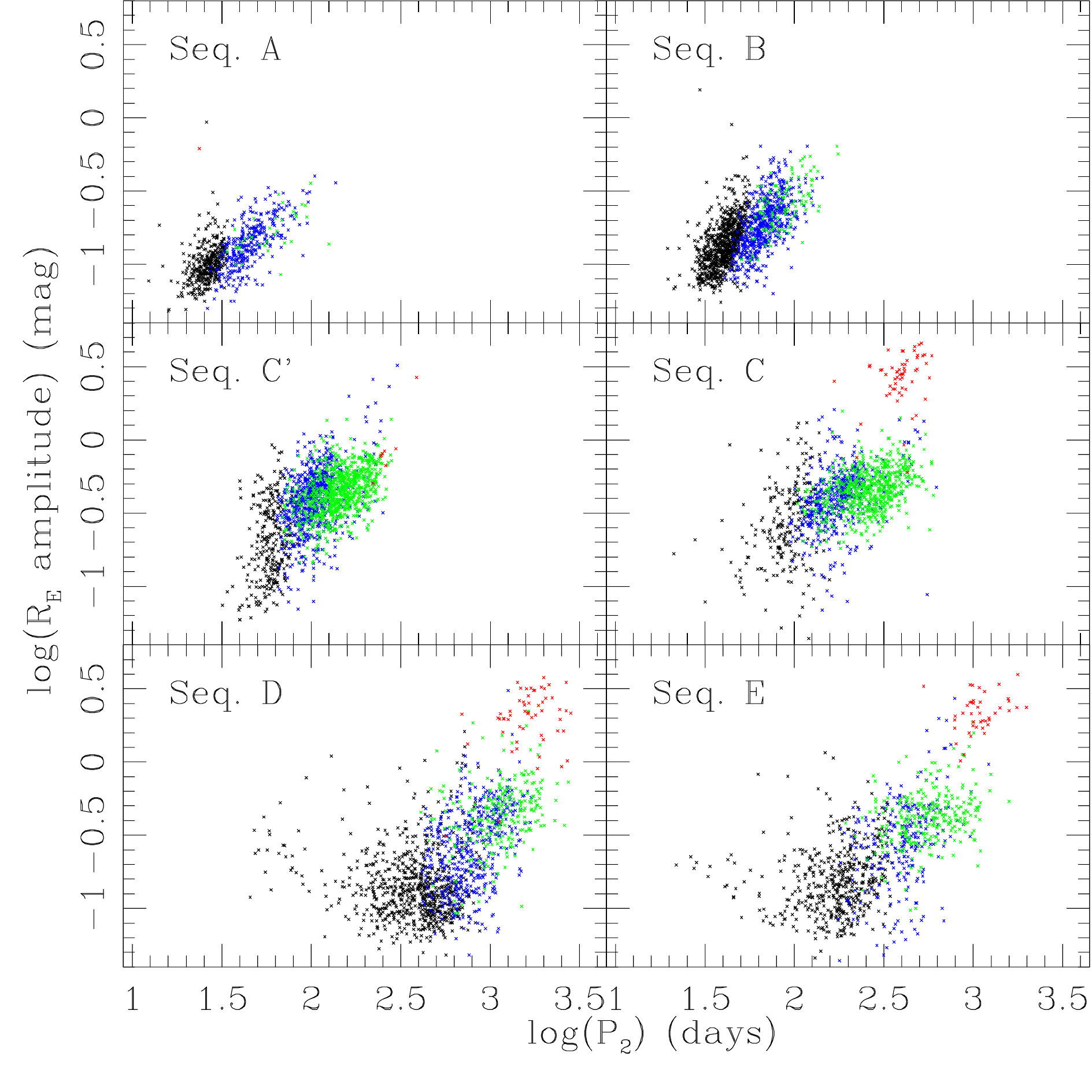}
   	\caption{Period-amplitude diagram, in log scale, for the second period of double-periodic LPVs according to the sequences of Fig.~\ref{3.6_logP_1Per}. }
   	\label{PerAmp2PerP2}
\end{figure}

Figures~\ref{PerAmp2PerP1} and \ref{PerAmp2PerP2} show the period-amplitude diagrams for, respectively, the first and second periods of double-periodic LPVs for each of the sequences defined in Sect.~\ref{SeqId}.
Overtone pulsators from sequences A and B display the same amplitude and period properties as for mono-periodic LPVs.

Most O-rich and C-rich variables from sequence C (excluding the extreme AGB stars) have amplitudes smaller than 1 magnitude, whereas mono-periodic variables extend to larger amplitudes. In addition they have the same amplitude range as the ones in sequence C'. This suggests that double-periodic LPVs from sequences C and C' are semi-regular variables with each of their periods in one of these sequences. Extreme AGB stars are still among the largest amplitude LPVs and hence are clearly distinguished from the other types. Most of them have one of their periods on sequence D (cf. Fig.~\ref{3.6_logP_2PerP2_P1seq}), which makes them LSP variables.

Except for RGB stars with the shortest periods, almost all double-periodic stars from sequences E and D follow period-amplitude relations, the shorter the period the smaller the amplitude. The RGB, O-rich, and C-rich stars have amplitudes smaller than one magnitude, and the differences between the mean amplitudes of the RGB stars, O-rich, and C-rich AGB stars are larger for these two sequences than for the other ones (cf. Table~\ref{table4}). Extreme AGB stars are LPVs with the longest periods and the largest amplitudes.

\section{Summary}
\label{Sect:Conclusions}
We have performed a systematic search for LPVs in the variable star database from the EROS-2 survey of the LMC. We have shown that applying the criterion for the Abbe value r of $r<0.4$ leads to a selection of smooth light curves over the 6.5 year duration of the survey. A cut based on the red giant clump color was used to separate our LPV candidates from other types of variable stars.\\
We have created a catalog of 43\,551 LPV candidates. Combining three period-search methods, we obtain periods for more than 86\% of these candidates. Thanks to data in $J$, $K_{s}$, and $[3.6]$ bands from 2MASS and \textit{Spitzer}, we have determined the O-rich or C-rich status of our LPV candidates. We retrieved the classical sequences in the period-magnitude diagram studied by \citet{Woodetal99} and \citet{Itaetal04}. We have demonstrated that these sequences are easier to distinguish using $[3.6]$ data than in the near infrared. The main improvement is for the extreme AGB stars that are enshrouded into a dust shell whose spectral energy distribution peaks around a few microns.\\
Our analysis of period-magnitude diagrams has shown that mono-periodic LPVs are preferentially distributed on the fundamental mode sequence C and ellipsoidal binaries from sequence E. Double-periodic LPVs have their first period mostly distributed in sequence D and their second period in first/second overtone sequences C' and B.
 Extreme AGB stars appear as the most luminous ones and are mainly found on the fundamental mode sequence C of the mono-periodic cases. The C-rich stars studied are visible from sequence C' up to the highest periods in sequence D. Less luminous O-rich AGB stars are distributed all along the sequences and are clearly separated from the RGB stars by a clear drop in the stellar density at $[3.6]=11.85$ mag. The RGB stars are spread over all the sequences and constitute the main component of sequence E, which consists of binary systems.
Our study of period-amplitude diagrams has shown that amplitudes tend to increase slowly as we consider longer periods from RGB to C-rich AGB stars. Extreme AGB stars have the highest amplitudes among LPVs in our sample, well above those of the other subtypes.
Our analysis of the complete EROS-2 variable stars database to extract LPVs remains ongoing work that will allow us to perform a homogeneous study of LPVs in different metallicity environments.

\begin{acknowledgements}
     We thank everyone from the EROS-2 collaboration for the access granted to the database.
     MS is grateful to Eric Lesquoy for his technical help and for his efforts allowing access to the database.
     The EROS-2 project was funded by the CEA and the CNRS through the IN2P3 and INSU institutes.\\
     This work was partly funded by Fonds national suisse de la recherche scientifique (FNRS).\\
    This publication makes use of data products from the Two Micron All Sky Survey, which is a joint project of the University of Massachusetts and the Infrared Processing and Analysis Center/California Institute of Technology, funded by the National Aeronautics and Space Administration and the National Science Foundation.\\
This work is based in part on observations made with the Spitzer Space Telescope, which is operated by the Jet Propulsion Laboratory, California Institute of Technology under a contract with NASA.\\
     We would like to thank Richard I. Anderson for his careful reading of the paper.
\end{acknowledgements}

\bibliography{biblio.bib}
\bibliographystyle{aa}

\end{document}